\documentclass[11pt]{article}
\usepackage{typearea}\typearea{13}
\usepackage{epsfig,amsmath,amsfonts,amssymb,makeidx,ifthen}

\newcommand{\url}[1]{{\tt #1}}

\makeatletter
\long\def\@makecaption#1#2{{\small
\advance\leftskip1cm
\advance\rightskip1cm
\vskip\abovecaptionskip
\sbox\@tempboxa{#1: #2}%
\ifdim \wd\@tempboxa >\hsize
 #1: #2\par
\else
\global \@minipagefalse
\hb@xt@\hsize{\hfil\box\@tempboxa\hfil}%
\fi
\vskip\belowcaptionskip}}
\makeatother

\newenvironment{FRAME}{\begin{trivlist}\item[]
\hrule
\hbox to \linewidth\bgroup
\advance\linewidth by -30pt
\hsize=\linewidth
\vrule\hfill
\vbox\bgroup
\vskip15pt

\begin{minipage}{\linewidth}}{%
\end{minipage}\vskip15pt
\egroup\hfill\vrule
\egroup\hrule
\end{trivlist}}

{\end{equation}\end{FRAME}}
{\end{align}\end{FRAME}}

\newcommand{\bigno}{\par\bigskip\noindent}

\newcommand{\nl}{\notag\\}

\makeatletter
\@addtoreset{equation}{section}
\makeatother

\def\eq#1\en{\begin{equation}#1\end{equation}}  
\def\eqsplit#1\ensplit{
	\begin{equation}\begin{split}#1\end{split}\end{equation}
	}
\def\eqalign#1\enalign{
	\begin{align}#1\end{align}
	}
\def\eqa#1\ena{
	\begin{align}#1\end{align}
	}
\def\eqg#1\eng{
	\begin{gather}#1\end{gather}
}
\def\eqmul#1\enmul{
	\begin{multline}#1\end{multline}
	}
\newcommand{\lb}[1]  {\label{e:#1}}
\newcommand{\rlb}[1] {\eqref{e:#1}}     
\newtheorem{theorem}{Theorem}[section]
\newtheorem{T}[theorem]{Theorem}
\newtheorem{C}[theorem]{Corollary}

\newtheorem{Le}[theorem]{Lemma}

\newcommand{\nproof}[1]{\par\noindent{\em #1\/}:}
\newcommand{\sproof}[1]{\par\bigskip\noindent{\em #1\/}:}
\newcommand{\qedm}{\rule{1.5mm}{3mm}}
\newcommand{\abs}[1]{\left|#1\right|}
\newcommand{\norm}[1]{\left\Vert#1\right\Vert}

\newcommand{\snorm}[1]{\Vert#1\Vert}
\newcommand{\rbk}[1]{\left(#1\right)}

\newcommand{\bkt}[1]{\left\langle#1\right\rangle}
\newcommand{\bbkt}[1]{\bigl\langle#1\bigr\rangle}

\newcommand{\sbkt}[1]{\langle#1\rangle}

\newcommand{\bra}[1]{\langle#1|}
\newcommand{\ket}[1]{|#1\rangle}

\newcommand{\sumtwo}[2]%
{\mathop{\sum_{#1}}_{#2}}
\newcommand{\sumthree}[3]%
{\mathop{\mathop{\sum_{#1}}_{#2}}_{#3}}
\newcommand{\sumfour}[4]%
{\mathop{\mathop{\mathop{\sum_{#1}}_{#2}}_{#3}}_{#4}} 
\newcommand{\prodtwo}[2]%
{\mathop{\prod_{#1}}_{#2}}
\newcommand{\mintwo}[2]%
{\mathop{\min_{#1}}_{#2}}
\newcommand{\maxtwo}[2]%
{\mathop{\max_{#1}}_{#2}}
\newcommand{\maxthree}[3]%
{\mathop{\mathop{\max_{#1}}_{#2}}_{#3}}
\newcommand{\limtwo}[2]%
{\mathop{\lim_{#1}}_{#2}}
\newcommand{\suptwo}[2]%
{\mathop{\sup_{#1}}_{#2}}
\newcommand{\supthree}[3]%
{\mathop{\mathop{\sup_{#1}}_{#2}}_{#3}}
\newcommand{\supfour}[4]%
{\mathop{\mathop{\mathop{\sup_{#1}}_{#2}}_{#3}}_{#4}} 
\newcommand{\inftwo}[2]%
{\mathop{\inf_{#1}}_{#2}}
\newcommand{\infthree}[3]%
{\mathop{\mathop{\inf_{#1}}_{#2}}_{#3}}
\newcommand{\inffour}[4]%
{\mathop{\mathop{\mathop{\inf_{#1}}_{#2}}_{#3}}_{#4}} 

\newcommand\calO{{\cal O}}

\newcommand{\bsn}{\boldsymbol{n}}

\newcommand{\hf}{\hat{f}}
\newcommand{\hg}{\hat{g}}
\newcommand{\hh}{\hat{h}}
\newcommand{\ho}{\hat{o}}

\newcommand{\hp}{\hat{p}}

\newcommand{\hA}{\hat{A}}

\newcommand{\hC}{\hat{C}}

\newcommand{\hH}{\hat{H}}

\newcommand{\bbR}{\mathbb{R}}
\newcommand{\bbZ}{\mathbb{Z}}

\newcommand{\up}{\uparrow}
\newcommand{\dn}{\downarrow}

{\end{itemize}\end{FRAME}}
\newcounter{mondaiCounter}[subsection]
\renewcommand{\themondaiCounter}{\thesubsection.\alph{mondaiCounter}}
\newcounter{mondaiCounterS}[section]
\renewcommand{\themondaiCounterS}{\thesection.\alph{mondaiCounterS}}
\newcommand{\toi}[1]{
\ifthenelse{\value{subsection}=0}
{\refstepcounter{mondaiCounterS}
\par\bigskip\noindent\underline{Problem \themondaiCounterS}\label{#1}~(\hyperlink{s:#1}{solution$\to$})~}
{\refstepcounter{mondaiCounter}
\par\bigskip\noindent\underline{Problem \themondaiCounter}\label{#1}~(\hyperlink{s:#1}{solution$\to$})~}
}
\newcommand{\toin}[1]{
\ifthenelse{\value{subsection}=0}
{\refstepcounter{mondaiCounterS}
\par\noindent\underline{Problem \themondaiCounterS}\label{#1}~(\hyperlink{s:#1}{solution$\to$})~}
{\refstepcounter{mondaiCounter}
\par\noindent\underline{Problem \themondaiCounter}\label{#1}~(\hyperlink{s:#1}{solution$\to$})~}
}

\newcounter{reiCounter}[subsection]
\renewcommand{\thereiCounter}{\thesubsection.\alph{reiCounter}}
\newcounter{reiCounterS}[section]
\renewcommand{\thereiCounterS}{\thesection.\alph{reiCounterS}}
\newcommand{\rei}[1]{
\ifthenelse{\value{subsection}=0}
{\refstepcounter{reiCounterS}
\par\bigskip\noindent{\bf Example \thereiCounterS}\label{#1}~}
{\refstepcounter{reiCounter}
\par\bigskip\noindent{\bf Exmaple \thereiCounter}\label{#1}~}
}

\newcommand{\hbS}{\hat{\boldsymbol{S}}}
\newcommand{\hSo}{\hat{S}^{(1)}}
\newcommand{\hSt}{\hat{S}^{(2)}}
\newcommand{\hSs}{\hat{S}^{(3)}}

\newcommand{\sumoL}{\sum_{x=1}^L}
\newcommand{\tenoL}{\bigotimes_{x=1}^L}

\newcommand{\La}{\Lambda}

\newcommand{\la}{\lambda}

\newcommand{\LaL}{\Lambda_L}

\newcommand{\Zd}{\bbZ^d}

\newcommand{\hOL}{\hat{\calO}_L}
\newcommand{\hOLa}{\hat{\calO}_L^{(\alpha)}}
\newcommand{\hOLo}{\hat{\calO}_L^{(1)}}
\newcommand{\hOLt}{\hat{\calO}_L^{(2)}}
\newcommand{\hOLs}{\hat{\calO}_L^{(3)}}

\newcommand{\hoa}{\ho^{(\alpha)}}
\newcommand{\hoo}{\ho^{(1)}}
\newcommand{\hot}{\ho^{(2)}}
\newcommand{\hos}{\ho^{(3)}}

\newcommand{\Phik}{\ket{\Phi}}
\newcommand{\Phiup}{\ket{\Phi_L^\up}}
\newcommand{\Phidn}{\ket{\Phi_L^\dn}}

\newcommand{\Gak}{\ket{\Gamma_L}}

\newcommand{\Xip}{\ket{\Xi_L^+}}

\newcommand{\GS}{\ket{\Phi_L^\mathrm{gs}}}
\newcommand{\GSb}{\bra{\Phi_L^\mathrm{gs}}}

\newcommand{\EGS}{E_L^{\rm gs}}
\newcommand{\Efst}{E_L^{\rm 1st}}
\newcommand{\MmL}{M_{\rm max}(L)}

\newcommand{\ms}{m^*}

\newcommand{\limM}{\lim_{M\up\infty}}

\newcommand{\limL}{\lim_{L\up\infty}}
\newcommand{\limn}{\lim_{n\up\infty}}
\newcommand{\limV}{\lim_{V\up\infty}}

\newcommand{\hHL}{\hat{H}_L}

\begin{document}
\renewcommand{\thefootnote}{\fnsymbol{footnote}}
\begin{flushright}
\footnotesize
Revised version, July 28, 2019.
\end{flushright}
\noindent
{\Large\bf Long-range order, ``tower'' of states, and symmetry breaking \\in lattice quantum systems\footnote{
Published in Journal of Statistical Physics {\bf 174}, 735--761 (2019).
This version is more complete than the published version.}}
\par\bigskip
\noindent
Hal Tasaki\footnote{
Department of Physics, Gakushuin University, Mejiro, Toshima-ku, 
Tokyo 171-8588, Japan
}
\renewcommand{\thefootnote}{\arabic{footnote}}
\setcounter{footnote}{0}
\begin{quotation}
\small
In a quantum many-body system where the Hamiltonian and the order operator do not commute, it often happens that the unique ground state of a finite system exhibits long-range order (LRO) but does not show spontaneous symmetry breaking (SSB).
Typical examples include antiferromagnetic quantum spin systems with N\'eel order, and lattice boson systems which exhibit Bose-Einstein condensation.
By extending and improving previous results by Horsch and von der Linden and by Koma and Tasaki, we here develop a fully rigorous and almost complete theory about the relation between LRO and SSB in the ground state of a finite system with continuous symmetry.
We show that a ground state with LRO but without SSB is inevitably accompanied by a series of energy eigenstates, known as the ``tower'' of states, which have extremely low excitation energies.
More importantly, we also prove that one gets a physically realistic ``ground state'' by taking a superposition of these low energy excited states.

The present paper is written in a self-contained manner, and does not require any knowledge about the previous works on the subject.
\end{quotation}

\tableofcontents
\section{Introduction}
\label{s:intro}

\begin{figure}
\centerline{\epsfig{file=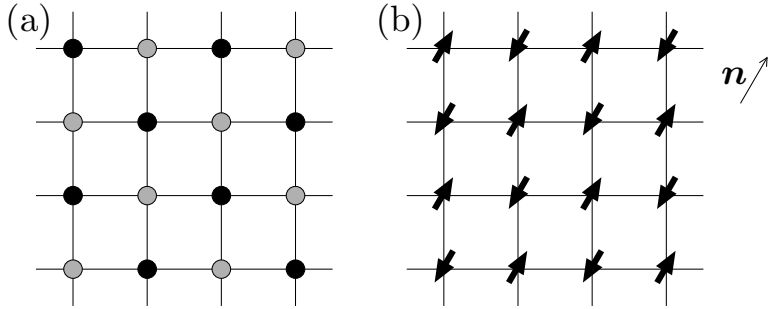,width=10cm}}
\caption[dummy]{
(a)~The square lattice, which is the $d$-dimensional hypercubic lattice with $d=2$, with $L=4$.
We impose periodic boundary conditions.
The lattice is connected and bipartite.
Sites in one sublattice are drawn in black and sites in the other in gray.

(b)~N\'eel order in the square lattice.
An arbitrary direction $\bsn$ is chosen (by the system).
Spins in one sublattice are pointing in the direction $\bsn$, and those in the other sublattice are pointing in the direction $-\bsn$.
This is an example of spontaneous symmetry breaking (SSB).
}
\label{f:ABN}
\end{figure}

\paragraph{Long-range order and Spontaneous symmetry breaking}
The antiferromagnetic Heisenberg model with Hamiltonian
\eq
\hHL=\frac{1}{2}\sumtwo{x,y\in\LaL}{(|x-y|=1)}\hbS_x\cdot\hbS_y,
\lb{HHAF}
\en
on the $d$-dimensional $L\times\cdots\times L$ hypercubic lattice $\LaL$ with even $L$ and periodic boundary conditions (see \rlb{MHA2} for the definition) exhibits antiferromagnetic  long-range order (LRO) in the ground state provided that $d\ge2$.
More precisely, it has been proved rigorously (except for the only case with $d=2$ and $S=1/2$) that the correlation function in the ground state behaves as
\eq
\GSb\hbS_x\cdot\hbS_y\GS\simeq
\begin{cases}
q_0&\text{if $x,y\in A$ or $x,y\in B$},\\
-q_0&\text{if $x\in A$, $y\in B$ or $x\in B$, $y\in A$},
\end{cases}
\lb{MHA1}
\en
with the long-range order parameter $q_0>0$, provided that the sites $x$ and $y$ are sufficiently far apart \cite{DysonLiebSimon1978,NevesPerez1986,KennedyLiebShastry1988a}.
Here $A$ and $B$ are sublattices with the property that any pair of neighboring sites belong to different sublattices.
See Fig.~\ref{f:ABN}~(a).
In short, two spins on the same sublattice tend to point in the same direction, while two spins in different sublattices tend to point in the opposite directions, no matter how separated the locations of two spins are.

The existence of long-range order suggests that the ground state also exhibits antiferromagnetic order (also called  N\'eel order), i.e., there is a preferred direction, say $\bsn$, and all the spins in the $A$ sublattice almost point in the direction $\bsn$, and all the spins in the $B$ sublattice almost point in the direction $-\bsn$.
See Figure~\ref{f:ABN}~(b).
In fact this is what is observed, through, e.g., neutron scattering experiments and NMR measurements, in actual quantum antiferromagnets at very low temperatures.
In a recent experiment with a system of cold atoms simulating the Heisenberg antiferromagnet, the ordering is observed much more directly \cite{Mazurenko2017}.

Note that, in this description of a N\'eel ordered ground state, a direction $\bsn$ should be specified.
Since the Hamiltonian \rlb{HHAF} is completely isotropic (or, equivalently, SU(2) invariant), the direction $\bsn$ must be chosen in an arbitrary manner by the system itself.
This arbitrary choice is known as spontaneous symmetry breaking (SSB).
In this case it is the SU(2) symmetry of the Hamiltonian \rlb{HHAF} that is spontaneously broken.

\paragraph{LRO without SSB}
It was proved by Marshall \cite{Marshall} and Lieb and Mattis \cite{LiebMattis1962} that the ground state of \rlb{HHAF} for any finite even $L$ is unique.
The uniqueness  is in sharp conflict with the idea of SSB.
The unique ground state must preserve all the symmetries of the Hamiltonian, and can never exhibit any order in a specific direction.
Therefore the unique ground state of the antiferromagnetic Heisenberg model exhibits LRO but no SSB.
Physically speaking, such a state is quite unnatural since a macroscopic quantity (i.e., the N\'eel order parameter)  then exhibits large fluctuation as in \rlb{QIsing9} and \rlb{LSHA0}, thus violating the law of large number.
Indeed, in magnetic systems, states with LRO without SSB are never observed experimentally; physically natural states with both LRO and SSB are observed.
Nature chooses to break symmetry rather than breaking the law of large number.

This puzzle about the discrepancy between physically natural ``ground states'' (with SSB) and the mathematical  ground state (without SSB) has been solved by now.
The solution, which will be carefully described here with the aid of mathematically rigorous results, consists of two essential observations.
First such a ground state with LRO but without SSB is inevitably accompanied by a series of low-lying energy eigenstates, which have very small excitation energies.\footnote{%
These energy eigenstates must not be confused with the spin-wave excitations.
See the remark at the end of section~\ref{ss:LSHA1}.
}
(See the remark below \rlb{LLS} for a precise definition of low-lying states.)
The number of such low energy eigenstates increases indefinitely as the system size grows.
The series of states is often called the ``tower of states'' or ``Anderson's tower of states''.
Secondly, the physical ``ground states'' with explicit SSB are not exact energy eigenstates, but are particular superpositions of the low energy excited states forming the ``tower''.
We stress that this picture applies universally to almost any quantum many-body systems exhibiting LRO in which the Hamiltonian and the order operator do not commute with each other.
Examples include antiferromagnetism, superconductivity\footnote{%
Here we mean the standard textbook treatment of superconductivity where dynamical electromagnetic field is not included.
}, Bose-Einstein condensate, and any quantum field theory (with LRO), but not ferromagnetism.

The above mentioned puzzle and its solution were probably realized (at least intuitively) since the early days of research on antiferromagnetism.
It was mentioned, e.g., in the seminal paper in 1952 by Anderson \cite{Anderson1952}, who seems to have understood the basic picture.
In the introduction of Anderson's book \cite{Anderson1984} published in 1984, there is a clear discussion about the role of the ``tower'' of energy eigenstates in the formation of N\'eel ordered states.
Anderson then continues as follows.
\begin{quote}
But somehow this is one of those arguments that is, although very simple, and terribly important, not generally available, perhaps because everyone who has ever understood it thinks it too simple to write down.  (p.~44, \cite{Anderson1984})
\end{quote}
In the present paper, we shall write down this fascinating and universal picture, emphasizing rigorous results which provide a firm basis for the picture.

\paragraph{Previous works and the present work}
The ``tower'' of  low-lying energy eigenstates and its relation to N\'eel ordered physical ``ground state'' were observed and discussed mainly in the context of numerical diagonalization of quantum spin systems.
Unlike in experiments, where a state similar to physical ``ground states'' should be observed, one directly observes exact ground states and low-lying energy eigenstates in such numerical studies \cite{Gross89,KikuchiOkabeMiyashita1990}.\footnote{%
It seems that people started observing the tower structure numerically in the early 90's when sufficiently advanced computers became available.
We find, for example, partial data for the tower in Table~I of \cite{Gross89}, and a complete tower structure in Table~I of \cite{KikuchiOkabeMiyashita1990}, both for the $S=1/2$ antiferromagnetic Heisenberg model on the square lattice.
}
The peculiar tower structure of the spectrum can be used as an indication for the existence of SSB in the physical ``ground state'' \cite{BernuLhuillierPierre92}.
See, e.g., \cite{AzariaDelamotteMouhanna1993,Lhuillier} for general pictures.

For quantum spin models where discrete symmetry is spontaneously broken, a general and mathematically rigorous theory about the relation between LRO and SSB was developed by  Horsch and von der Linden \cite{HorschLinden}.
They proved that  a ground state with LRO but without SSB must be accompanied by a low-lying excited state.
But in these models there appear only a finite number of low-lying energy eigenstates; there is no ``tower'' of states.
It was also proved by Kaplan, Horsch, and von der Linden \cite{KaplanHorschLinden} that the ground state of the infinite system exhibits SSB when infinitessimally small symmetry breaking field is applied.

Koma and Tasaki studied quantum systems on a lattice where the relevant symmetry is continuous and the unique ground state exhibits LRO but not SSB \cite{KomaTasaki1993,KomaTasaki1994}.
It was proved that there inevitably appears an ever increasing number of low-lying excited states;
the existence of the ``tower of states'' was established rigorously \cite{KomaTasaki1994}.
It was also proved that one can construct a low energy state which explicitly breaks the symmetry by superposing these low-lying excited states \cite{KomaTasaki1993}.
Koma and Tasaki conjectured that this symmetry breaking state is the physically relevant ``ground state'' which exhibits small fluctuation \cite{KomaTasaki1994}.

\bigskip

In this paper, we present an almost complete rigorous theory about the relation between LRO and SSB, which considerably improves that of Koma and Tasaki.
Our Theorem~\ref{t:LLS}, which establishes the existence of low-lying excited states, is a strict extension of the theorem by Koma and Tasaki \cite{KomaTasaki1994}.
It was assumed in \cite{KomaTasaki1994} that the model possesses $U(1)\times\bbZ_2$ symmetry, while we only assume $U(1)$ symmetry.
This extension makes it possible to apply the theorem to lattice boson systems other than at half-filling.
We also note that our proof is much simpler than that in \cite{KomaTasaki1994}.

The most important contribution of the present paper is Theorem~\ref{t:main} about the symmetry breaking state obtained by summing up a series of low-energy eigenstates as in  \cite{KomaTasaki1993}.
It was shown in \cite{KomaTasaki1993} that the state breaks the symmetry of the model to the full extent, but we here also prove that the order operator has vanishing fluctuation in the infinite volume limit of this state.
We have thus confirmed the conjecture by Koma and Tasaki that this state is a physically meaningful ``ground state'' which shows SSB.

We here describe a general theory, but keeping quantum antiferromagnetic system in mind.
Another important class of models to which our theory readily applies  is that of systems of hard core bosons on a lattice, idealized models of ultra cold atoms in an optical lattice.
See, e.g., \cite{BlochDalibardZwerger}.
Although these systems are mathematically almost the same as quantum antiferromagnets, the physical interpretation of LRO and SSB is rather different and needs to be discussed carefully.
We shall leave detailed discussion to \cite{TasakiBOOK}.
See also \cite{TasakiBoson}.


\bigskip
The present paper is organized as follows.
In section~\ref{s:QIsing}, we focus on the very elementary example of one-dimensional quantum Ising model to illustrate the nature of a ground state with LRO but without SSB.
We also discuss the general theories of Horsch and von der Linden  and of Kaplan, Horsch, and von der Linden in this context.
Then, in section~\ref{s:LSHA}, we introduce the class of models that we study, and describe all the rigorous results in the general setting.
All the theorems are proved in section~\ref{s:proof}.
We conclude the paper by discussing some open problems in section~\ref{s:discussion}.

\section{Quantum Ising model}
\label{s:QIsing}
To motivate the main topic of the paper, we shall briefly study the simplest model where one encounters a ground state with LRO but without SSB, namely, the quantum Ising model.
After discussing the nature of low energy eigenstates of the model, we see how the general theories of Horsch and von der Linden \cite{HorschLinden} and Kaplan, Horsch, and von der Linden \cite{KaplanHorschLinden} apply to the system.

\subsection{Low energy eigenstates and physical ``ground states''}
Consider the one-dimensional quantum Ising model with $S=1/2$, whose Hamiltonian is
\eq
\hHL=-\sum_{x=1}^{L}\hSs_x\hSs_{x+1}-\lambda\sumoL\hSo_x,
\lb{QIsing1}
\en
where we write the spin operator at site $x=1,\ldots,L$ as $\hbS_x=(\hSo_x,\hSt_x,\hSs_x)$, and take the periodic boundary condition $\hbS_{L+1}=\hbS_1$.
The first term precisely corresponds to the classical ferromagnetic Ising model, and the second term, which describes the external magnetic field with magnitude $\lambda\ge0$ in the 1-direction, introduces quantum nature to the model.
Note that this field is not a symmetry breaking field.
The Hamiltonian is invariant under the $\pi$ rotation about the 1-axis, $(\hSo_x,\hSt_x,\hSs_x)\to(\hSo_x,-\hSt_x,-\hSs_x)$.
The symmetry group is $\bbZ_2$.
To detect possible symmetry breaking we introduce the order operator  
\eq
\hOL=\sumoL\hSs_x.
\lb{QIsing1B}
\en

When $\lambda=0$, the problem is trivial.
There are two degenerate ground states
\eq
\Phiup:=\tenoL\ket{\!\up}_x,\quad
\Phidn:=\tenoL\ket{\!\dn}_x,
\lb{QIsing2}
\en
which are accompanied by a finite energy gap equal to 1.
Both $\Phiup$ and $\Phidn$ exhibit symmetry breaking as well as LRO.

\begin{figure}
\centerline{\epsfig{file=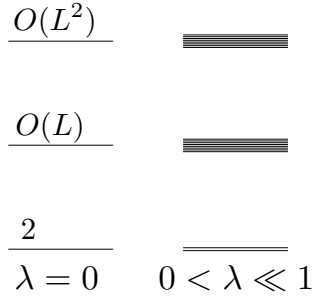,width=4cm}}
\caption[dummy]{
A schematic picture of the low-energy spectra of the Hamiltonian \rlb{QIsing1} for $\lambda=0$ and for $0<\lambda\ll1$.
In the classical Ising model with $\lambda=0$, the energy levels are separated by $1$.
The ground states are doubly degenerate, and the degeneracies of the  second and the third levels are of $O(L)$ and $O(L^2)$, respectively.
When the perturbation $0<\lambda\ll1$ is turned on, the degeneracies are lifted while the overall structure of the spectrum is unchanged.
}
\label{f:QIsing}
\end{figure}

When $0<\lambda\ll1$, the degeneracy of the ground states is lifted.
See Figure~\ref{f:QIsing}.
A perturbative analysis shows that the model \rlb{QIsing1} has a unique ground state
\eq
\GS\simeq\frac{1}{\sqrt{2}}\bigl(\Phiup+\Phidn\bigr),
\lb{QIsing5}
\en
and the first excited state
\eq
\ket{\Phi_L^{\rm 1st}}\simeq\frac{1}{\sqrt{2}}\bigl(\Phiup-\Phidn\bigr),
\lb{QIsing6}
\en
whose energy eigenvalue is denoted as $\Efst$.
It is also found that the energy difference between the ground state and the first excited state is exponentially small in the system size, i.e., $\Efst-\EGS\simeq (\text{const})\lambda^{L}$.
Note that both $\GS$ and $\ket{\Phi_L^{\rm 1st}}$ are Schr\"{o}dinger's cat-like states, in which two macroscopically distinct states are superposed.

The unique ground state \rlb{QIsing5} clearly exhibits LRO as
\eq
\GSb\Bigl(\frac{\hOL}{L}\Bigr)^2\GS
\simeq\frac{1}{2}\Bigl\{
\bra{\Phi_L^\up}\Bigl(\frac{\hOL}{L}\Bigr)^2\Phiup+
\bra{\Phi_L^\dn}\Bigl(\frac{\hOL}{L}\Bigr)^2\Phidn
\Bigr\}
=\frac{1}{4},
\lb{QIsing7}
\en
but no SSB as
\eq
\GSb\Bigl(\frac{\hOL}{L}\Bigr)\GS
\simeq\frac{1}{2}\Bigl\{
\bra{\Phi_L^\up}\Bigl(\frac{\hOL}{L}\Bigr)\Phiup+
\bra{\Phi_L^\dn}\Bigl(\frac{\hOL}{L}\Bigr)\Phidn
\Bigr\}=0.
\lb{QIsing8}
\en
This means that the density $\hOL/L$ shows nonvanishing fluctuation in this ground state:
\eq
\sqrt{
\GSb\Bigl(\frac{\hOL}{L}\Bigr)^2\GS
-
\Bigl\{\GSb\Bigl(\frac{\hOL}{L}\Bigr)\GS\Bigr\}^2
}
\simeq\frac{1}{2}
\lb{QIsing9}
\en
Recall that, in physically realistic states, any macroscopic quantity obeys the law of large number, in the sense that it shows vanishing fluctuation in the macroscopic limit.
Although being the exact unique ground state, the state $\GS$, in which the law of large number is violated, cannot be regarded as a physical state of a macroscopic system.

The solution to this ``puzzle'' should be clear.
The all-up state $\Phiup$ and the all-down state $\Phidn$ (with inevitable small modifications) are the physically meaningful ``ground states'' which are relevant to experimental observations in macroscopic systems.
These states are not eigenstates of the Hamiltonian for any finite $L$, but a physical picture strongly suggest that they represent realistic  ``ground states'' that we expect to observe.
We call them physical ``ground states''.\footnote{%
The quotation marks indicated that they are not ground states in the standard definition in quantum mechanics.
}

We point out that this trivial example provides some hints for general theories.
First, from \rlb{QIsing5} and \rlb{QIsing6}, one finds
\eq
\Phiup\simeq\frac{1}{\sqrt{2}}\bigl(\GS+\ket{\Phi_L^{\rm 1st}}\bigr),\quad
\Phidn\simeq\frac{1}{\sqrt{2}}\bigl(\GS-\ket{\Phi_L^{\rm 1st}}\bigr),
\lb{QIsing10}
\en
which suggests that physical ``ground states'' may in general be obtained as liner combinations of the exact ground state and a low-lying energy eigenstate.
Secondly, by noting that $\hOL\Phiup=(L/2)\Phiup$ and $\hOL\Phidn=-(L/2)\Phidn$, one sees from \rlb{QIsing5} and \rlb{QIsing6} that
\eq
\hOL\GS\simeq\hOL\frac{1}{\sqrt{2}}\bigl(\Phiup+\Phidn\bigr)
\simeq\frac{L}{2\sqrt{2}}\bigl(\Phiup-\Phidn\bigr)
\simeq\frac{L}{2}\ket{\Phi_L^{\rm 1st}}.
\lb{QIsing11}
\en
This suggests that a low-lying energy eigenstate may be constructed (at least approximately) by operating the order operator onto the exact ground state.
Finally, by combining \rlb{QIsing10} and \rlb{QIsing11}, we find for the present example that
\eq
\Phiup\simeq\frac{1}{\sqrt{2}}\Bigl(\GS+\frac{2}{L}\hOL\GS\Bigr),\quad
\Phidn\simeq\frac{1}{\sqrt{2}}\Bigl(\GS-\frac{2}{L}\hOL\GS\Bigr).
\lb{QIsing12}
\en
These relations are quite interesting since the physical ``ground states'' are constructed entirely out of the exact ground state $\GS$.
We will show that analogous construction works in much more general settings.
See \rlb{Gen10} and \rlb{LSHA5}.

\subsection{General theory of low-lying states and SSB}
\label{s:Gen}
We shall review the theories of  Horsch and von der Linden \cite{HorschLinden} and  Kaplan, Horsch, and von der Linden \cite{KaplanHorschLinden}.
Although we treat the quantum Ising model \rlb{QIsing1} for simplicity, it should be clear that the arguments apply to more general models in higher dimensions.

The only properties we need about the ground state $\GS$ is that it exhibits LRO but not SSB in the sense that 
\eqg
\GSb\Bigl(\frac{\hOL}{L}\Bigr)^2\GS\ge q_0,
\lb{Gen2}\\
\GSb\Bigl(\frac{\hOL}{L}\Bigr)^n\GS=0\ \text{for $n=1,3$},
\lb{Gen3}
\eng
with nonzero long-range order parameter $q_0$.
(We have $q_0\simeq1/4$, but it is only essential that $q_0>0$.)

Suggested by \rlb{QIsing11}, we introduce the normalized trial state of Horsch and von der Linden \cite{HorschLinden}:
\eq
\Gak=\frac{\hOL\GS}{\snorm{\hOL\GS}}
\lb{Gen4}
\en
Note that the condition \rlb{Gen3} implies the orthogonality $\GSb\Gamma\rangle=0$.
By recalling $\snorm{\hOL\GS}^2=\GSb(\hOL)^2\GS$, we estimate the energy expectation value of the trial state $\Gak$ as
\eqa
\langle\Gamma_L|\hHL|\Gamma_L\rangle-\EGS
&=\frac{
\GSb\hOL\hHL\hOL\GS-\EGS\GSb(\hOL)^2\GS
}{\GSb(\hOL)^2\GS}
\nl
&=\frac{
\GSb\hOL\hHL\hOL\GS-\frac{1}{2}\GSb(\hOL)^2\hHL\GS-\frac{1}{2}\GSb\hHL(\hOL)^2\GS
}{\GSb(\hOL)^2\GS}
\nl
&=\frac{\GSb[\hOL,[\hHL,\hOL]]\GS}{2\GSb(\hOL)^2\GS}.
\lb{Gen5}
\ena
Write the Hamiltonian \rlb{QIsing1} as $\hHL=\sumoL\hh_x$ with $\hh_x=-\hSs_x\hSs_{x+1}-\lambda\hSo_x$, and also write $\ho_x=\hSs_x$.
Then we can rewrite the double commutator in the right-hand side of \rlb{Gen5} as
\eq
[\hOL,[\hHL,\hOL]]=
\sumoL\bigl\{
[\ho_x,[\hh_x,\ho_x]]+[\ho_{x+1},[\hh_x,\ho_x]]+[\ho_x,[\hh_x,\ho_{x+1}]]+[\ho_{x+1},[\hh_x,\ho_{x+1}]]
\bigr\}
\lb{Gen7}
\en
By using the  simple norm estimate\footnote{%
We denote by $\snorm{\hA}$ the standard operator norm defined as $\snorm{\hA}:=\sup\snorm{\hA\Phik}$, where the supremum is taken over all $\Phik$ such that $\snorm{\Phik}=1$.
} $\snorm{[\ho_x,[\hh_y,\ho_z]]}\le 4\snorm{\ho_x}\,\snorm{\hh_y}\,\snorm{\ho_z}\le \{(1/4)+\lambda\}$, we find 
\eq
\GSb[\hOL,[\hHL,\hOL]]\GS\le \Bigl(\frac{1}{4}+\lambda\Bigr)L.
\lb{Gen8}
\en
Substituting this bound and the assumption \rlb{Gen2} about LRO into \rlb{Gen5}, we get
\eq
0\le\langle\Gamma_L|\hHL|\Gamma_L\rangle-\EGS\le\frac{C}{L},
\lb{Gen9}
\en
with the constant $C=(1+4\lambda)/(8q_0)$.
We thus see that $\Gak$ is a  low-lying state, i.e., a state (not necessarily an energy eigenstate) whose energy expectation value approaches the ground state energy as $L$ becomes large.

The existence of a low-lying state which is orthogonal to the ground state implies, from the standard  variational argument, the existence of an energy eigenstate  $\ket{\Psi_L}$ such that $\GSb\Psi_L\rangle=0$ whose energy eigenvalue $E$ satisfies $\EGS\le E\le\EGS+C/L$.
Thus $\ket{\Psi_L}$ is a low-lying excited state.
Note that the  upper bound $C/L$ of the energy gap is much larger than the exponentially small energy gap $\Efst-\EGS\simeq(\text{const})\lambda^{L}$.

\paragraph{Low-lying states with LRO and SSB}
The next step of the general theory is to construct a low-lying state which exhibits both LRO and SSB.

Again, suggested by the observation \rlb{QIsing12}, we define another trial state
\eq
\Xip=\frac{1}{\sqrt{2}}\bigl(\GS+\Gak\bigr),
\lb{Gen10}
\en
which obviously satisfies $\langle\Xi_L^+\Xip=1$ and $\langle\Xi_L^+|\hHL\Xip\le\EGS+(C/2)L^{-1}$.
The state $\Xip$ is a low-lying state.

Recalling the definition \rlb{Gen4} of $\Gak$, we evaluate the expectation value of the order parameter as
\eqa
\langle\Xi_L^+|\hOL\Xip&=
\frac{1}{2}\biggl\{
\biggl(\GSb+\frac{\GSb\hOL}{\snorm{\hOL\GS}}\Bigr)
\hOL
\Bigl(\GS+\frac{\hOL\GS}{\snorm{\hOL\GS}}\biggr)
\biggr\}
\nl
&=\frac{\GSb(\hOL)^2\GS}{\snorm{\hOL\GS}}
=\sqrt{\GSb(\hOL)^2\GS},
\lb{Gen11}
\ena
where we used the assumption \rlb{Gen3} about the absence of SSB in $\GS$.
With the assumption \rlb{Gen2} about the existence of LRO, this means
\eq
\langle\Xi_L^+|\frac{\hOL}{L}\Xip\ge\sqrt{q_0}.
\lb{Gen12}
\en
We thus conclude that $\Xip$ is a low-lying state which exhibits both LRO and SSB.
Of course the state $\ket{\Xi_L^-}=(\GS-\Gak)/\sqrt{2}$, which is orthogonal to $\ket{\Xi_L^+}$, is a low-lying state that satisfies $\bra{\Xi_L^-}(\hOL/L)\ket{\Xi_L^-}\le-\sqrt{q_0}$.

For the quantum Ising model and similar models with discrete symmetry breaking, the low-lying states $\Xip$ and $\ket{\Xi_L^-}$ are indeed physical ``ground states'' in which the fluctuation of the order parameter vanishes as $L\up\infty$.
This fact has been proved rigorously in Appendix~B of \cite{KomaTasaki1994}.
We find that, in the antiferromagnetic Heisenberg model (and other models with continuous symmetry), the states $\Xip$ and $\ket{\Xi_L^-}$ are still pathological states in the sense that the density of the order operator shows huge fluctuation.

\paragraph{SSB under infinitesimal symmetry breaking field}
We shall finally show that SSB can be triggered by infinitesimal symmetry breaking field.
The low-lying state $\Xip$ plays an essential role here.

The following elegant variational argument is due to Kaplan, Horsch, and von der Linden \cite{KaplanHorschLinden}.
Define the Hamiltonian with symmetry breaking field $h\ge0$ as
\eq
\hHL^h=\hHL-h\hOL,
\lb{Gen14}
\en
where $\hHL$ is the original Hamiltonian in \rlb{QIsing1}, and $h$ is the standard magnetic filed since  $\hOL=\sumoL\hSs_x$.

Let $\ket{\Phi_L^{{\rm gs},h}}$ be a ground state of the new Hamiltonian $\hHL^h$.
Although we don't know almost anything about $\ket{\Phi_L^{{\rm gs},h}}$, we at least know from the definition of ground state that
\eq
\bra{\Xi_L^+}\hHL^h\Xip\ge\bra{\Phi_L^{{\rm gs},h}}\hHL^h\ket{\Phi_L^{{\rm gs},h}}.
\lb{Gen15}
\en
By substituting \rlb{Gen14}, and arranging the inequality, we find
\eqa
\bra{\Phi_L^{{\rm gs},h}}\frac{\hOL}{L}\ket{\Phi_L^{{\rm gs},h}}
&\ge
\langle\Xi_L^+|\frac{\hOL}{L}\Xip
+\frac{1}{hL}\bigl\{
\bra{\Phi_L^{{\rm gs},h}}\hHL\ket{\Phi_L^{{\rm gs},h}}-\langle\Xi_L^+|\hHL\Xip
\bigr\}
\nl&
\ge\sqrt{q_0}+\frac{1}{hL}\bigl\{
\EGS-\langle\Xi_L^+|\hHL\Xip
\bigr\},
\lb{Gen16}
\ena
where we used the essential bound \rlb{Gen12} and a trivial inequality $\bra{\Phi_L^{{\rm gs},h}}\hHL\ket{\Phi_L^{{\rm gs},h}}\ge\EGS$ to get the second inequality.
We know that $0\ge\EGS-\langle\Xi_L^+|\hHL\Xip\ge-(C/2)L^{-1}$, and hence the second term in the right-most-hand vanishes as $L\up\infty$.
Thus the following theorem was proved \cite{KaplanHorschLinden}.
\begin{T}[Kaplan-Horsch-von der Linden theorem]
\label{t:KHL}
Assume \rlb{Gen2} and  \rlb{Gen3}.
Then we have
\eq
\lim_{h\dn0}\limL
\bra{\Phi_L^{{\rm gs},h}}\frac{\hOL}{L}\ket{\Phi_L^{{\rm gs},h}}
\ge\sqrt{q_0}.
\lb{Gen17}
\en
\end{T}
Note that the order of the limits is essential, since one obviously has 
\newline$\lim_{h\dn0}\bra{\Phi_L^{{\rm gs},h}}({\hOL}/{L})\ket{\Phi_L^{{\rm gs},h}}=0$ by continuity.

\section{Models with continuous symmetry}
\label{s:LSHA}

Let us now turn to the main topic of the present paper and study quantum spin systems in which continuous symmetry is spontaneously broken.
A typical and most important example is the antiferromagnetic Heisenberg model \rlb{HHAF} discussed in section~\ref{s:intro}.

A natural strategy here is to follow the observation and the general theory in section~\ref{s:QIsing} to construct physical ``ground states'' with both LRO and SSB.
However it is clear from the outset that the framework in section~\ref{s:QIsing} is not enough to solve the problem.
Here we expect N\'eel order in which an arbitrary direction $\bsn$ should be specified (see section~\ref{s:intro}), or, equivalently, a spontaneous breakdown of continuous SU(2) symmetry.
Then there should be infinitely many physical ``ground states'' corresponding to the infinitely many choices of the direction $\bsn$.
The construction in section~\ref{s:QIsing}, which essentially makes use of the ground state $\GS$ and a single low-lying state $\hOL\GS$, cannot give rise to infinitely many states with SSB.

In the present section, we discuss our rigorous results about low-lying states and symmetry breaking for systems with U(1) or SU(2) symmetry.
As we have stated in the introduction, our results are extensions and/or improvements of those by  Koma and Tasaki \cite{KomaTasaki1993,KomaTasaki1994}.
All the theorems are proved in section~\ref{s:proof}.

\subsection{General setting and assumptions}
Let us introduce the general class of models we study, and state precise assumptions.
For $d=1,2,\ldots$ and even $L$, let 
\eq
\LaL:=\bigl\{(x_1,x_2,\ldots,x_d)\,\bigl|\,x_i\in\bbZ,\ -\frac{L}{2}<x_i\le\frac{L}{2}\,\bigr\}\subset\Zd,
\lb{MHA2}
\en
be the $d$-dimensional hypercubic lattice with $V=L^d$ sites.
We consider a quantum many-body system on $\LaL$ with the general Hamiltonian 
\eq
\hHL=\sum_{x\in\LaL}\hh_x.
\en
We here impose periodic boundary conditions, and assume that the Hamiltonian has translation invariance in the sense that all $\hh_x$ are translational copies of each other.\footnote{%
Although we do not make use of the translation invariance in the proofs of the theorems, it is necessary to guarantee that a series of low-lying states converges to an infinite volume ground state.
See the remark after \rlb{LLS}.
}
We further assume that the model possesses the following properties:
\par\bigno
(A1)~There is a self-adjoint operator $\hC_L$ such that $[\hC_L,\hh_x]=0$ for any $x$, and hence $[\hC_L,\hHL]=0$.
This means that $\hC_L$ is a conserved charge.
We also assume that $-aV\le\hC_L\le aV$ with a constant $a>0$.
\par\bigno
(A2)~There are order operators $\hOLo=\sum_{x\in\LaL}\hoo_x$ and $\hOLt=\sum_{x\in\LaL}\hot_x$.
We assume that $\hoo_x$ and $\hot_x$ are self-adjoint and translationally invariant.
We also assume that the order operators satisfy the commutation relations $[\hOLo,\hC_L]=-i\hOLt$, $[\hOLt,\hC_L]=i\hOLo$, and $[\hOLo,\hOLt]=i\hC_L$.\footnote{%
The commutation relation $[\hOLo,\hOLt]=i\hC_L$ makes our discussion considerably simple, but may not be necessary.
We expect that one can prove basically the same results (with much more effort) without assuming it.
}

\par\bigno
(A3) The operator $\hh_x$ acts at most on $\zeta$ sites including $x$. 
The operators $\hoo_x$ and $\hot_x$ act only on the site $x$.\footnote{%
This condition about the support of $\hoo_x$ is introduced to make the proofs, especially that of Lemmas~\ref{l:R1} and \ref{l:R2}, easy.
One can extend the theory to cover $\hoo_x$ acting on more than one site, by constructing slightly more complicated inductive proof.
}
We write $h_0:=\snorm{\hh_x}$ and $o_0:=\snorm{\hoo_x\pm i\hot_x}$, where the norm is independent of $x$.
\par\bigno
(A4) The ground state $\GS$ of $\hHL$ for finite $L$ is unique and satisfies $\hC_L\GS=M_L\GS$.
\par\bigno
(A5) The ground state exhibits LRO in the sense that there is a constant $q_0$ independent of $L$, and
we have $\GSb(\hOLo/V)^2\GS=\GSb(\hOLt/V)^2\GS\ge q_0>0$.

\par\bigskip\par
From (A2) and (A4) it follows that $\GSb\hOLa\GS=0$ for $\alpha=1,2$.\footnote{%
The commutation relations in (A2) implies $e^{i\theta\hC_L}\hOLo e^{-i\theta\hC_L}=\cos\theta\,\hOLo-\sin\theta\,\hOLt$, and in particular $e^{i\pi\hC_L}\hOLo e^{-i\pi\hC_L}=-\hOLo$.
Then from (A4) one finds $\GSb\hOLo\GS=\GSb e^{i\pi\hC_L}\hOLo e^{-i\pi\hC_L}\GS=-\GSb\hOLo\GS$.
}
Then (A5) implies 
\eq
\sqrt{
\GSb\Bigl(\frac{\hOLa}{V}\Bigr)^2\GS
-
\Bigl\{\GSb\Bigl(\frac{\hOLa}{V}\Bigr)\GS\Bigr\}^2
}
\ge\sqrt{q_0},
\lb{LSHA0}
\en
for $\alpha=1,2$, from which find, exactly as in \rlb{QIsing9}, that the order operator $\hOLa$ violates the law of large number.

From  (A2) one sees that $(\hOLo,\hOLt,\hC_L)$ form generators of SU(2).
But we only require the model and the ground state to have U(1) symmetry as in (A1), (A4), and (A5).
Clearly the  operators
\eq
\hOL^\pm:=\hOLo\pm i\hOLt
\lb{LSHA1}
\en
satisfy the commutation relations $[\hC_L,\hOL^\pm]=\pm\hOL^\pm$, and act as raising and lowering operators, respectively, of the conserved quantity $\hC_L$.

\subsection{Examples}
One can easily write down various quantum models which satisfy the assumptions (A1), (A2), and (A3).
The uniqueness (A4) is not trivial, but still can be proved in a large class of models by using the technique developed by Lieb and Mattis \cite{LiebMattis1962}.
The assumption (A5), the existence of LRO, is the most important and nontrivial condition, which is usually very difficult to justify rigorously.
For the moment the existence of LRO associated with spontaneous breakdown of continuous symmetry can be proved only by using the reflection positivity method developed by Dyson, Lieb, and Simon \cite{DysonLiebSimon1978}.
Of course one can always apply our theorems to find the properties of low-lying excited states and symmetry breaking ``ground states'' when it is plausible that the model exhibits LRO in its ground state.

A typical example which satisfies the assumptions is the quantum XXZ model with general spin $S$, whose Hamiltonian is
\eq
\hHL=\frac{1}{2}\sumtwo{x,y\in\LaL}{(|x-y|=1)}\{\hSo_x\hSo_y+\hSt_x\hSt_y+\lambda\hSs_x\hSs_y\},
\lb{LSHA24}
\en
where we again denoted the spin operator at site $x\in\LaL$ as $\hbS_x=(\hSo_x,\hSt_x,\hSs_x)$.
We assume that $L$ is even and imposed periodic boundary conditions.
Here $\lambda$ is the Ising anisotropy parameter, and the model reduces to the antiferromagnetic Heisenberg model \rlb{HHAF} when $\lambda=1$.
Assumptions (A1), (A2), and (A3) are obviously satisfied if we set
\eq
\hOLo=\sum_{x\in\LaL}(-1)^x\hSo_x, \quad
\hOLt=\sum_{x\in\LaL}(-1)^x\hSt_x,\quad
\hC_L=\sum_{x\in\LaL}\hSs_x,
\en
where $(-1)^x=1$ if $x$ is in the sublattice $A$, and $(-1)^x=-1$ if $x$ is in $B$.
See Figure~\ref{f:ABN}~(a).

The uniqueness of the ground state of the model \rlb{LSHA24} was proved by Marshall \cite{Marshall} and by Lieb and Mattis \cite{LiebMattis1962} when $\lambda=1$.
The proof was extended to other values of $\lambda$ in \cite{Mattis79,Nishimori81}.
It is known that the ground state $\GS$ for a finite $L$ is unique and satisfies $\hC_L\GS=0$ provided that $\lambda>-1$.
Thus our assumption (A4) is satisfied with $M_L=0$.
Since we are interested in the breakdown of continuous symmetry, we do not consider the region with $\la>1$, where the model exhibits (or is expected to exhibit) Ising type long range order and/or symmetry breaking.

By developing a method based on the notion of reflection positivity,  Dyson, Lieb, and Simon proved the existence of LRO for the antiferromagnetic Heisenberg model with $d\ge3$ at sufficiently low temperatures \cite{DysonLiebSimon1978}.
The proof was later extended to cover the ground state and the anisotropic model \rlb{LSHA24}.
Now the existence of LRO in the ground state of \rlb{LSHA24}, which is the assumption (A5), is rigorously known for the model in $d=3$ with any $\lambda\in[0,1]$ and $S=1/2,1,\ldots$, and the model in $d=2$ with $\lambda\in[0,0.20)$ if $S=1/2$ and with $\lambda\in[0,1]$ if $S=1$ or larger.
See \cite{NevesPerez1986,KennedyLiebShastry1988a,KennedyLiebShastry1988b,KuboKishi1988,OzekiNishimoriTomita1989} and references therein.

Our theory also applies to the system of hard core bosons, which is mathematically equivalent to the  XXZ model with $S=1/2$.
In this model it is expected that the ground state exhibits LRO even when $M_L/L\ne0$, to which case Theorem~\ref{t:LLS} below applies.
But the existence of LRO is proved only for $M_L=0$.
See \cite{TasakiBOOK} for discussion about the difference and similarity between the system of bosons and the quantum antiferromagnets.

\subsection{The ``tower'' of low-lying energy eigenstates}
\label{ss:LSHA1}
We first discuss the result which establishes the existence of the ``tower'' of states.
By using  the  ``raising'' and ``lowering'' order operators \rlb{LSHA1}, we define a series of trial states
\eq
\ket{\Gamma_L^M}=\frac{(\hOL^+)^M\GS}{\snorm{(\hOL^+)^M\GS}},\quad
\ket{\Gamma_L^{-M}}=\frac{(\hOL^-)^M\GS}{\snorm{(\hOL^-)^M\GS}},
\lb{LSHA2}
\en
for $M=1,2,\ldots$,
which should be compared with $\Gak$ defined in \rlb{Gen4}.
Since $\hC_L\GS=M_L\GS$, we see $\hC_L\ket{\Gamma_L^M}=(M_L+M)\ket{\Gamma_L^M}$.
This in particular means that all $\ket{\Gamma_L^M}$ are orthogonal with each other.
We then have the following theorem.
\begin{T}
\label{t:LLS}
There are constants\footnote{%
The constants depend only on $o_0$, $h_0$, $\zeta$, and $q_0$.
See the proof, for example \rlb{C1choice}, for explicit dependences.}
$C_1$, $C_2$, and $C_3$.
For any $M$ such that $|M|\le C_1\sqrt{V}$, the state $\ket{\Gamma_L^M}$ is well-defined, and satisfies
\eq
\bra{\Gamma_L^M}\hHL\ket{\Gamma_L^M}\le\EGS+C_2\frac{M^2}{V},
\lb{LSHA3}
\en
if $M_L=0$\footnote{%
The condition $M_L=0$ can be replaced by the condition that $|M_L|$ is bounded by a constant independent of $L$.
}, and
\eq
\bra{\Gamma_L^M}\hHL\ket{\Gamma_L^M}\le\EGS+C_3\frac{M^3}{V},
\lb{LSHA3new}
\en
in general.
\end{T}
The bound \rlb{LSHA3} was proved by Koma and Tasaki \cite{KomaTasaki1994} by assuming the higher $\mathrm{U}(1)\times\bbZ_2$, rather than the U(1) symmetry.
The bound  \rlb{LSHA3new} for the general case is new.
In fact we strongly expect that the bound of the form  \rlb{LSHA3} is valid for the general case as well, but we still cannot prove it.

Theorem~\ref{t:LLS} implies that for any $M$ such that $|M|\le C_1\sqrt{V}$, we have
\eq
\frac{1}{V}\bigl\{\bra{\Gamma_L^M}\hHL\ket{\Gamma_L^M}-\EGS\bigr\}\le
\begin{cases}
(C_1)^2C_2\,V^{-1}&\text{if $M_L=0$},\\
(C_1)^3C_3\,V^{-1/2}&\text{if $M_L\ne0$}.
\end{cases}
\lb{LLS}
\en
Note that the right-hand side converges to zero as $L\up\infty$.
We shall call a state $\ket{\Gamma_L^M}$ with such a property a low-lying state since its energy density converges to that of the ground state.\footnote{
To be more precise it is crucial that the excitation is spread almost uniformly over the whole lattice in the state $\ket{\Gamma_L^M}$.
A state obtained by, for example, exciting a single spin from the ground state satisfies a similar bound as \rlb{LLS}, but it is regarded as an excited state.
}
It is known that if a series of low-lying states converges to a state of the infinite system, the limiting state is an infinite volume ground state.
See \cite{KomaTasaki1994,TasakiBOOK} for details.

The existence of low-lying states immediately implies the existence of low-lying energy eigenstates.
\begin{C}
For any $M$ such that $|M|\le C_1\sqrt{V}$, there exists an energy eigenstate $\ket{\Psi_L^M}$ such that $\hC_L\ket{\Psi_L^M}=(M_L+M)\ket{\Psi_L^M}$ whose energy eigenvalue $E_L^M$ satisfies
\eq
\EGS<E_L^M\le\EGS+C_2\frac{M^2}{V},
\lb{LSHA4}
\en
if $M_L=0$ and
\eq
\EGS<E_L^M\le\EGS+C_3\frac{M^3}{V},
\lb{LSHA4new}
\en
in general.
\end{C}
Remarkably we have established that there are $O(\sqrt{V})$ distinct low-lying energy eigenstates in the system with linear size $L$.
Comparing this with the simple energy spectrum of the quantum Ising model, where the ground state and a single low-lying energy eigenstate were separated from other energy eigenstates by a finite gap (see Figure~\ref{f:QIsing}), one sees that the existence of ever increasing number of low-lying energy eigenstates (which form the ``tower'') is a clear manifestation of LRO without SSB related to a continuous symmetry. 

An approximate mean field type theory for the ``tower'' of states \cite{AzariaDelamotteMouhanna1993,Lhuillier} predicts that the low-lying energy eigenvalues are given by 
\eq
E_L^M-\EGS\simeq(\text{constant})\,\frac{M(M+1)}{V}.
\lb{LSHA5B}
\en
This formula in fact fits numerical data quite well.
It is also interesting that the scaling $M^2/V$ in the rigorous upper bound  \rlb{LSHA4} basically recovers this behavior.
This is in contrast to the case of quantum Ising model, where the actual energy gap is $O(\lambda^{L})$ while the upper bound is $O(L^{-1})$.
See section~\ref{s:QIsing}.

The low-lying excited states discussed here should not be confused with spin wave excitations (or, equivalently, the Nambu-Goldston mode) in quantum antiferromagnets, whose excitation energies are proportional to $L^{-1}$ rather than $V^{-1}=L^{-d}$.
The states with spin waves should be obtained by modifying the ``ground state'' which has explicit antiferromagnetic order.
We may say that the spin wave excitations are relevant to actual experimental observations, while the low-lying excitations are relevant to the exact energy spectrum of a large but finite systems.
Recently Koma has constructed a series of low energy excitations above the infinite volume ground state with explicit SSB in the antiferromagnetic Heisenberg model \cite{Koma2018}.
As far as we know this is the first rigorous demonstration of the existence of spin wave excitations in quantum antiferromagnets.
See references in \cite{Koma2018} for background.

\subsection{Symmetry breaking}
\label{ss:LSHA2}
We next discuss results about the ``ground state'' with explicit symmetry breaking.

\paragraph{Low-lying states with full symmetry breaking}
We shall construct a low-lying state which breaks the symmetry to the full extent.
The construction is similar to that of $\ket{\Xi_\pm}$ in section~\ref{s:Gen}, but we can here make use of the ever increasing number of low-lying states in the  ``tower'' of states.

Let us define the symmetry breaking order parameter for the ground state by
\eq
\ms:=\lim_{k\up\infty}\limL\Bigl\{
\GSb\Bigl(\frac{\hOLa}{V}\Bigr)^{2k}\GS
\Bigr\}^{1/(2k)},
\lb{LSHA6}
\en
where $k$ takes integral values, and $\alpha=1$ or 2.
The existence of the $k\up\infty$ limit is guaranteed by Lemma~\ref{L:LSP2}.\footnote{%
The existence of the limit $L\up\infty$, on the other hand, is not guaranteed in general (although it is very much expected).
One may take a subsequence or replace $\lim$ with $\limsup$ or $\liminf$ if necessary.
}
As we see below in Theorem~\ref{t:KomaTasaki3}, that $q_0>0$ guarantees $\ms>0$.
Roughly speaking, $\ms$ is the maximum value that $|\hOLa/V|$ can take in the infinite volume limit.\footnote{%
The order of the limits is essential here.
If one takes the limit $k\up\infty$ for finite $L$ one simply gets the maximum possible value $S$, which does not reflect the properties of the ground state.
}

For an integer valued function $\MmL>0$ such that $\MmL\le C_1\sqrt{V}$, we define a trial state by summing up low-lying states as
\eq
\ket{\Xi_L^{(1,0)}}=\frac{1}{\sqrt{2\MmL+1}}\sum_{M=-\MmL}^{\MmL}
\ket{\Gamma_L^M},
\lb{LSHA5}
\en
where we  identified $\ket{\Gamma_L^0}$ with $\GS$.
Since this is a linear combination of low-lying states, we have
\eq
\lim_{L\up\infty}\frac{1}{V}\bigl\{\langle\Xi_L^{(1,0)}|\hH_L|\Xi_L^{(1,0)}\rangle-\EGS\bigr\}=0,
\en
which means that $\ket{\Xi_L^{(1,0)}}$ converges to an infinite volume ground state.

The following theorem, which is the most important contribution of the present work, shows that this trial state exhibits full symmetry breaking.\footnote{%
Note on June 23, 2019: 
In the previous versions of the paper, we noted that \rlb{LSHA9} and \rlb{LSHA10} with $\hOLt$ replaced by $\hOLs$ are valid in models with SU(2) symmetry.
This is true for \rlb{LSHA9}, but we still do not have a proof of the relation corresponding to \rlb{LSHA10}.}
\begin{T}
\label{t:main}
If $\MmL$ diverges to infinity not too rapidly as $L\up\infty$, one has
\eqg
\limL\bra{\Xi_L^{(1,0)}}\Bigl(\frac{\hOLo}{V}\Bigr)\ket{\Xi_L^{(1,0)}}=\ms,
\lb{LSHA7}\\
\limL\bra{\Xi_L^{(1,0)}}\Bigl(\frac{\hOLo}{V}\Bigr)^2\ket{\Xi_L^{(1,0)}}=(\ms)^2,
\lb{LSHA8}\\
\bra{\Xi_L^{(1,0)}}\Bigl(\frac{\hOLt}{V}\Bigr)\ket{\Xi_L^{(1,0)}}=0,
\lb{LSHA9}\\
\limL\bra{\Xi_L^{(1,0)}}\Bigl(\frac{\hOLt}{V}\Bigr)^2\ket{\Xi_L^{(1,0)}}=0.
\lb{LSHA10}
\eng
\end{T}

The inequality (which shows that the left-hand side is not smaller than the right-hand side) corresponding to \rlb{LSHA7} was proved as (7.26) by Koma and Tasaki in \cite{KomaTasaki1993}.
The equality  \rlb{LSHA7}, as well as \rlb{LSHA8} and \rlb{LSHA10} is new (while \rlb{LSHA9} is trivial).

Note first that \rlb{LSHA7} and \rlb{LSHA9} show that the state $\ket{\Xi_L^{(1,0)}}$ exhibits symmetry breaking in which the order operator (viewed as a two component vector) is pointing in the $(1,0)$ direction.
It is essential here that, in the limit $L\up\infty$, the expectation value of  $\hOLo/V$ is exactly equal to the order parameter $\ms$, which is designed to pick up the maximum value of $|\hOLa/V|$.
This means that the state $\ket{\Xi_L^{(1,0)}}$ breaks the symmetry to the full extent.

The full symmetry breaking also manifests in the expectation values of $(\hOLa/V)^2$, especially in that of $(\hOLo/V)^2$.
That we got $(\ms)^2$ in the right-hand side of \rlb{LSHA8} is an indication that we are here dealing with a macroscopically ``healthy'' state.
In particular  \rlb{LSHA7}, \rlb{LSHA8}, \rlb{LSHA9}, and \rlb{LSHA10} together imply that
\eq
\limL\sqrt{
\bra{\Xi_L^{(1,0)}}\Bigl(\frac{\hOLa}{V}\Bigr)^2\ket{\Xi_L^{(1,0)}}
-
\Bigl(\bra{\Xi_L^{(1,0)}}\frac{\hOLa}{V}\ket{\Xi_L^{(1,0)}}\Bigr)^2
}=0,
\lb{LSHA11}
\en
for $\alpha=1$ and 2, which means that the fluctuation of the order operator density $\hOLa/V$ vanishes in the limit $L\up\infty$.
We conclude that the low-lying state $\ket{\Xi_L^{(1,0)}}$, which exhibits LRO and full SSB, is the desired physical ``ground state'', as was conjectured by Koma and Tasaki \cite{KomaTasaki1994}.
(But see section~\ref{s:discussion}.)

\paragraph{Long-range order parameter and symmetry breaking order parameter}
So far we have defined two different order parameters, the long-range order parameter $q_0$ which characterizes the expectation value of $(\hOLa/V)^2$ as in the assumption (A5), and the symmetry breaking order parameter $\ms$ which is the maximum possible value of $|\hOLa/V|$ defined as in \rlb{LSHA6}.
These two are related by the following inequalities proved by Koma and Tasaki \cite{KomaTasaki1993}.
\begin{T}
\label{t:KomaTasaki3}
The two order parameters satisfy
\eq
\ms\ge\sqrt{2q_0},
\lb{LSHA20new}
\en
in general, and 
\eq
\ms\ge\sqrt{3q_0},
\lb{LSHA20}
\en
when the model has SU(2) symmetry.
\end{T}
We say that a model has SU(2) symmetry if the order operators $(\hOLo,\hOLt,\hOLs)$ transform as a vector under any three-dimensional rotation, and also the ground state is invariant under an arbitrary rotation.

The factor $\sqrt{2}$ or $\sqrt{3}$, which was absent in the corresponding relation \rlb{Gen12} for the quantum Ising model, reflects the U(1) or SU(2) symmetry of the model.
For the XXZ model \rlb{LSHA24}, the stronger inequality \rlb{LSHA20} applies to the case with $\lambda=1$, and \rlb{LSHA20new} to the case with other $\lambda$.

Although the above theorem is not new, we here present a complete proof, which is much simpler than the original proof in \cite{KomaTasaki1993}.
In particular the inequality \rlb{LSHA20new} for models with U(1) symmetry is derived in an almost trivial manner.

The appearance of the  factor $\sqrt{2}$ or $\sqrt{3}$ is not difficult to understand, at least intuitively.
Consider the case with U(1) symmetry.
Theorem~\ref{t:main} suggests that $\hat{\boldsymbol{o}}=(\hOLo/V,\hOLt/V)$ basically behaves as a classical vector of magnitude $\ms$ in physical ``ground states'' when $L$ is large.
In the state $\ket{\Xi_L^{(1,0)}}$, for example, one has $\hat{\boldsymbol{o}}\simeq(\ms,0)$.
In the unique ground state, which is rotationally symmetric, the behavior of $\hat{\boldsymbol{o}}$ is far from that of a classical vector, but we assume that the magnitude $\ms$ may be computed from the expectation value of  $(\hat{\boldsymbol{o}})^2$.
This leads to the estimate
\eq
(\ms)^2\simeq\GSb(\hat{\boldsymbol{o}})^2\GS=\sum_{\alpha=1}^2\GSb\Bigl(\frac{\hOLa}{V}\Bigr)^2\GS\ge2q_0,
\lb{LSHA21}
\en
where we used the assumption (A5).

\paragraph{SSB under infinitesimal symmetry breaking field}
It is obvious that the theorem of Kaplan, Horsch, and von der Linden, Theorem~\ref{t:KHL}, applies to the present general models.

We define the Hamiltonian with symmetry breaking field $h\ge0$, which is the staggered magnetic field in the XXZ model \rlb{LSHA24}, as
\eq
\hHL^h=\hHL-h\hOLo,
\lb{LSHA22}
\en
and let $\ket{\Phi_L^{{\rm gs},h}}$ be its (not necessarily unique) ground state.
Repeating the variational proof by using $\ket{\Xi_L^{(1,0)}}$ as a trial state, we get the following.
\begin{T}
\label{t:KHL2}
We have
\eq
\lim_{h\dn0}\limL
\bra{\Phi_L^{{\rm gs},h}}\frac{\hOLo}{V}\ket{\Phi_L^{{\rm gs},h}}
\ge\ms.
\lb{LSHA23}
\en
\end{T}

\section{Proofs}
\label{s:proof}
We shall describe the proofs of all the theorems.
In section~\ref{s:op} we introduce the new order operator \rlb{LSp2}, which enables us to make full use of the U(1) symmetry, and to extend the results in the previous works \cite{KomaTasaki1993,KomaTasaki1994}.
Then  we prove Theorems~\ref{t:main}, which is our main result about the symmetry breaking ``ground state'', in section~\ref{s:sb}, and Theorem~\ref{t:KomaTasaki3} in the (difficult) case with SU(2) symmetry in section~\ref{s:su2}.
These proofs are not too heavy, and we hope that the reader may find some arguments interesting.

On the other hand, the proof of Theorems~\ref{t:LLS} is rather involved (although it is much simpler than the complicated proof in \cite{KomaTasaki1994} of a more restricted theorem).
We shall carefully describe the proof in section~\ref{s:lls}.

Throughout the present section, we abbreviate the ground state expectation $\GSb\cdots\GS$ as $\sbkt{\cdots}$.

\subsection{Order parameters}
\label{s:op}
We use operators per unit volume
\eq
\hoa:=\frac{\hOLa}{V},\quad\ho^\pm:=\frac{\hOL^\pm}{V}=\hoo\pm i\hot,
\lb{LSp1}
\en
where $V=L^d$ is the volume.
We introduce the new order operator
\eq
\hp:=\frac{1}{2}(\ho^+\ho^-+\ho^-\ho^+)=(\hoo)^2+(\hot)^2,
\lb{LSp2}
\en
which turns out to be extremely useful.

Let us write $\ho^\pm_x=\hoo_x\pm i\hot_x$.
Noting that $\ho^\pm_x$ acts locally on site $x$, we have 
\eq
[\hOL^+,\hOL^-]=\sum_{x,y\in\La}[\ho_x^+,\ho_y^-]=\sum_{x\in\La}[\ho_x^+,\ho_x^-].
\lb{LSp3}
\en
This implies that
\eq
\norm{[\ho^+,\ho^-]}=\frac{1}{V^2}\norm{[\hOL^+,\hOL^-]}\le\frac{2(o_0)^2}{V},
\lb{LSp4}
\en
where $o_0$ is introduced in the assumption (A3).
This immediately implies the following elementary lemma.
\begin{Le}
\label{L:LSP1}
Let $s_1,\ldots,s_{2n}=\pm$ be any sequence such that $\sum_{j=1}^{2n}s_j=0$.
Then we have
\eq
\norm{\ho^{s_1}\ldots\ho^{s_{2n}}-\hp^n}\le\frac{2n^2(o_0)^{2n}}{V}.
\lb{LSp5}
\en
\end{Le}
\nproof{Proof}
Note that $\ho^{s_1}\ldots\ho^{s_{2n}}$ can be rearranged to any other $\ho^{s'_1}\ldots\ho^{s'_{2n}}$ with $\sum_{j=1}^{2n}s'_j=0$ by making at most $n^2$ exchanges of neighboring $\ho^+$ and $\ho^-$.
Then \rlb{LSp4} implies \rlb{LSp5}.~\qedm

\bigskip

For a positive integer $n$, the Schwarz inequality implies
\eq
\sbkt{\hp^n}^2=\sbkt{\hp^{(n-1)/2}\hp^{(n+1)/2}}^2\le\sbkt{\hp^{n-1}}\sbkt{\hp^{n+1}},
\lb{LSp6}
\en
from which we find
\eq
\frac{\sbkt{\hp^n}}{\sbkt{\hp^{n-1}}}\le\frac{\sbkt{\hp^{n+1}}}{\sbkt{\hp^n}}.
\lb{LSp7}
\en
Recall that, from the assumption (A5), we have $\sbkt{\hp}=\sbkt{(\hoo)^2}+\sbkt{(\hot)^2}\ge 2q_0>0$.
Thus \rlb{LSp7} implies
\eq
\frac{\sbkt{\hp^n}}{\sbkt{\hp^{n-1}}}\ge 2q_0,
\lb{LSp8}
\en
and hence $\sbkt{\hp^n}\ge(2q_0)^n$.

The following lemma shows the essential property of the ratio ${\sbkt{\hp^n}}/{\sbkt{\hp^{n-1}}}$.
\begin{Le}
\label{L:LSP2}
We have
\eq
\ms=\limn\limL\sqrt{\frac{\sbkt{\hp^n}}{\sbkt{\hp^{n-1}}}},
\lb{LSp9}
\en
where $\ms$ is defined in \rlb{LSHA6}.
The $n\up\infty$ limit in \rlb{LSp9}and the $k\up\infty$ limit in \rlb{LSHA6}  exist.
\end{Le}

Note that \rlb{LSp9} along with \rlb{LSp8} implies that
\eq
\ms\ge\sqrt{2q_0}.
\lb{LSp9B}
\en
This is nothing but the first inequality \rlb{LSHA20new} of Theorem~\ref{t:KomaTasaki3}.

\sproof{Proof of Lemma~\ref{L:LSP2}}
The limit $n\up\infty$ in \rlb{LSp9} exists because of the monotonicity \rlb{LSp7} and the boundedness $\sbkt{\hp^n}/\sbkt{\hp^{n-1}}\le\snorm{\hp}$.
We first observe that  \rlb{LSHA6} is written as
\eq
\ms=\limn\limL\sqrt{\frac{\bkt{(\hoo)^{2n}}}{\bkt{(\hoo)^{2(n-1)}}}},
\lb{LSp10}
\en
which is easy to prove.
Let $r_n=\limL\sbkt{(\hoo)^{2n}}$ and $\mu^*=\lim_{n\up\infty}\sqrt{r_n/r_{n-1}}$.
The limit $n\up\infty$ again exists by monotonicity.
We wish to show $\mu^*=\ms$.
For any $\varepsilon>0$ there exists, by definition, $n_0$ such that $|\sqrt{r_n/r_{n-1}}-\mu^*|\le\varepsilon$ for any $n\ge n_0$.
Let $n\ge n_0$.
Noting that $\sqrt{r_n}=\sqrt{r_{n_0}}\prod_{k=n_0+1}^n\sqrt{r_k/r_{k-1}}$, we have
\eq
\sqrt{r_{n_0}}\,(\mu^*-\varepsilon)^{n-n_0}\le\sqrt{r_n}\le\sqrt{r_{n_0}}\,(\mu^*+\varepsilon)^{n-n_0},
\en
and hence
\eq
(r_{n_0})^{1/(2n)}\,(\mu^*-\varepsilon)^{1-(n_0/n)}\le(r_n)^{1/(2n)}\le(r_{n_0})^{1/(2n)}\,(\mu^*+\varepsilon)^{1-(n_0/n)}.
\en
Since $\ms=\lim_{n\up\infty}(r_n)^{1/(2n)}$, we find by letting $n\up\infty$ that $\mu^*-\varepsilon\le\ms\le\mu^*+\varepsilon$.
Since $\varepsilon>0$ is arbitrary we find $\mu^*=\ms$.

A less trivial part is to show that the limits in the right-hand sides of \rlb{LSp9} and  \rlb{LSp10} coincide.
Let us write $\hoo=(\ho^++\ho^-)/2$, and observe that
\eq
\bbkt{(\hoo)^{2n}}=\frac{1}{2^{2n}}\sumtwo{s_1,\ldots,s_{2n}=\pm}{(\sum_js_j=0)}
\sbkt{\ho^{s_1}\cdots\ho^{s_{2n}}},
\lb{LSp12}
\en
where we used the assumption (A4).
There are $(2n)!/(n!)^2$ terms in the sum in the right-hand side of \rlb{LSp12}.
From \rlb{LSp5}, we see that each term in the sum is equal to $\sbkt{\hp^n}+O(1/V)$.
We then find
\eq
\bbkt{(\hoo)^{2n}}=\frac{1}{2^{2n}}\,\frac{(2n)!}{(n!)^2}\,\sbkt{\hp^n}+O\Bigl(\frac{1}{V}\Bigr)
=\frac{(2n-1)!!}{(2n)!!}\,\sbkt{\hp^n}+O\Bigl(\frac{1}{V}\Bigr),
\lb{LSp13}
\en
which implies
\eq
\frac{\bbkt{(\hoo)^{2n}}}{\bbkt{(\hoo)^{2(n-1)}}}=\frac{2n-1}{2n}\,\frac{\sbkt{\hp^n}}{\sbkt{\hp^{n-1}}}+O\Bigl(\frac{1}{V}\Bigr).
\lb{LSp14}
\en
We thus see
\eq
\limn\limL\sqrt{\frac{\bbkt{(\hoo)^{2n}}}{\bbkt{(\hoo)^{2(n-1)}}}}=\limn\limL\sqrt{\frac{\sbkt{\hp^n}}{\sbkt{\hp^{n-1}}}},
\lb{LSp15}
\en
which proves the desired \rlb{LSp9}.~\qedm

\subsection{Symmetry breaking state}
\label{s:sb}
We are now ready to prove Theorem~\ref{t:main}, which shows that  $\ket{\Xi_L^{(1,0)}}$ defined in \rlb{LSHA5} by summing up the low-lying states is a physically natural state with full SSB.
For $M=1,2,\ldots$, we define 
\eq
\ket{\Xi^{(M)}_L}=\frac{1}{\sqrt{2M+1}}\biggl(
\GS+\sum_{n=1}^M\frac{(\ho^+)^n\GS}{\norm{(\ho^+)^n\GS}}
+\sum_{n=1}^M\frac{(\ho^-)^n\GS}{\norm{(\ho^-)^n\GS}}
\biggr).
\lb{LSp20}
\en
Note that $\ket{\Xi_L^{(1,0)}}=\ket{\Xi^{(\MmL)}_L}$.

Multiplying \rlb{LSp20} by $\ho^+$, and organizing the terms, we have
\eq
\ho^+\ket{\Xi^{(M)}_L}=\frac{1}{\sqrt{2M+1}}\biggl(
\sum_{n=1}^{M+1}\frac{(\ho^+)^n\GS}{\norm{(\ho^+)^{n-1}\GS}}
+\sum_{n=1}^M\frac{\ho^+(\ho^-)^n\GS}{\norm{(\ho^-)^n\GS}}
\biggr).
\lb{LSp21}
\en
Rewriting \rlb{LSp20} as
\eq
\ket{\Xi^{(M)}_L}=\frac{1}{\sqrt{2M+1}}\biggl(
\sum_{n=1}^M\frac{(\ho^+)^n\GS}{\norm{(\ho^+)^n\GS}}
+\sum_{n=1}^{M+1}\frac{(\ho^-)^{n-1}\GS}{\norm{(\ho^-)^{n-1}\GS}}
\biggr),
\lb{LSp22}
\en
and using \rlb{LSp21}, we immediately get
\eqa
\bra{\Xi^{(M)}_L}\ho^+\ket{\Xi^{(M)}_L}&=\frac{1}{2M+1}\biggl(
\sum_{n=1}^M\frac{\bkt{(\ho^-)^n(\ho^+)^n}}{\norm{(\ho^+)^n\GS}\,\norm{(\ho^+)^{n-1}\GS}}
+
\sum_{n=1}^M\frac{\bkt{(\ho^+)^n(\ho^-)^n}}{\norm{(\ho^-)^{n-1}\GS}\,\norm{(\ho^-)^n\GS}}
\biggr)
\nl
&=\frac{1}{2M+1}\sum_{n=1}^M\biggl(
\sqrt{\frac{\bkt{(\ho^-)^n(\ho^+)^n}}{\bkt{(\ho^-)^{n-1}(\ho^+)^{n-1}}}}
+
\sqrt{\frac{\bkt{(\ho^+)^n(\ho^-)^n}}{\bkt{(\ho^+)^{n-1}(\ho^-)^{n-1}}}}
\biggr).
\lb{LSp23}
\ena
Note that there is symmetry $\bra{\Xi^{(M)}_L}\ho^+\ket{\Xi^{(M)}_L}=\bra{\Xi^{(M)}_L}\ho^-\ket{\Xi^{(M)}_L}$, which implies $\bra{\Xi^{(M)}_L}\hoo\ket{\Xi^{(M)}_L}=\bra{\Xi^{(M)}_L}\ho^+\ket{\Xi^{(M)}_L}$ and
\eq
\bra{\Xi^{(M)}_L}\hot\ket{\Xi^{(M)}_L}=0.
\lb{SP23BB}
\en
Thus the desired $\bra{\Xi^{(M)}_L}\hoo\ket{\Xi^{(M)}_L}$ is given by \rlb{LSp23}.
Again by using the  (controlled) approximation \rlb{LSp5} in the right-hand side of \rlb{LSp23}, we get
\eq
\bra{\Xi^{(M)}_L}\hoo\ket{\Xi^{(M)}_L}=\frac{2}{2M+1}\sum_{n=1}^M\sqrt{\frac{\bkt{\hp^n}}{\bkt{\hp^{n-1}}}}+O\Bigl(\frac{1}{V}\Bigr).
\lb{LSp24}
\en
We then find from \rlb{LSp9} that
\eq
\limM\limL\bra{\Xi^{(M)}_L}\hoo\ket{\Xi^{(M)}_L}=\ms.
\lb{LSp25}
\en

Let us evaluate the expectation value of $(\hoo)^2+(\hot)^2$.
Noting that this is nothing but $\hp$, we easily see from \rlb{LSp20} and   \rlb{LSp5} that
\eqa
\bra{\Xi^{(M)}_L}\{(\hoo)^2&+(\hot)^2\}\ket{\Xi^{(M)}_L}=\bra{\Xi^{(M)}_L}\hp\ket{\Xi^{(M)}_L}
\nl
&=\frac{1}{2M+1}\biggl(
\bkt{\hp}+\sum_{n=1}^M\frac{\bkt{(\ho^-)^n\hp(\ho^+)^n}}{\bkt{(\ho^-)^{n}(\ho^+)^{n}}}
+\sum_{n=1}^M\frac{\bkt{(\ho^+)^n\hp(\ho^-)^n}}{\bkt{(\ho^+)^{n}(\ho^-)^{n}}}
\biggr).
\nl&=\frac{1}{2M+1}\biggl(\bkt{\hp}+2\sum_{n=1}^M\frac{\bkt{\hp^{n+1}}}{\bkt{\hp^n}}
\biggr)+O\Bigl(\frac{1}{V}\Bigr).
\lb{LSp26}
\ena
This, with \rlb{LSp9}, implies that
\eq
\limM\limL\bra{\Xi^{(M)}_L}\{(\hoo)^2+(\hot)^2\}\ket{\Xi^{(M)}_L}=(\ms)^2.
\lb{LSp27}
\en
We see from \rlb{LSp25} and \rlb{LSp27} that
\eqa
(\ms)^2&=\limM\limL\bigl(\bra{\Xi^{(M)}_L}\hoo\ket{\Xi^{(M)}_L}\bigr)^2\le\limM\limL\bra{\Xi^{(M)}_L}(\hoo)^2\ket{\Xi^{(M)}_L}
\nl&\le\limM\limL\bra{\Xi^{(M)}_L}\{(\hoo)^2+(\hot)^2\}\ket{\Xi^{(M)}_L}=(\ms)^2,
\lb{LSp28}
\ena
which implies 
\eqg
\limM\limL\bra{\Xi^{(M)}_L}(\hoo)^2\ket{\Xi^{(M)}_L}=(\ms)^2,
\lb{LSp29A}
\\
\limM\limL\bra{\Xi^{(M)}_L}(\hot)^2\ket{\Xi^{(M)}_L}=0.
\lb{LSp29}
\eng

Now note that the relations  \rlb{LSp25}, \rlb{LSp29A},  \rlb{SP23BB}, and \rlb{LSp29}
precisely correspond to the desired relations \rlb{LSHA7},  \rlb{LSHA8},  \rlb{LSHA9}, and  \rlb{LSHA10}, respectively, except that we have the double limit $\limM\limL$ instead of a single limit $\limL$ where $M=\MmL$ varies according to $L$.
Intuitively speaking, the double limit  $\limM\limL$ corresponds to a single limit $\limL$ with $\MmL$ that diverges indefinitely slowly.
It only remains to extend the relations to $\MmL$ that does not diverge too rapidly, but this is only technical.

Let $\epsilon(n)>0$ be an arbitrary decreasing sequence such that $\epsilon(n)\dn0$ as $n\up\infty$.
For each $n$ we choose a positive integer $L(n)$ such that
\eq
\Biggl|
\frac{\GSb(\hp_L)^{k}\GS}{\GSb(\hp_L)^{k-1}\GS}
-
\lim_{L'\up\infty}\frac{\bra{\Phi_{\rm gs}^{L'}}(\hp_{L'})^{k}\ket{\Phi_{\rm gs}^{L'}}}{\bra{\Phi_{\rm gs}^{L'}}(\hp_{L'})^{k-1}\ket\Phi_{\rm gs}^{L'}}
\Biggr|
\le\epsilon(n)
\lb{LSp30newA}
\en
holds for any $k\le n$ whenever $L\ge L(n)$.
(We have made the size dependence of $\hp$ explicit.)
We can choose $L(n)$ to be increasing and to satisfy $L(n)\up\infty$ as $n\up\infty$.
For each $L$, we let $\MmL$ be the largest $k$ such that $L\ge L(k)$.
Clearly we have $\MmL\up\infty$ as $L\up\infty$.

For a given $L$, take any $k$ such that $k\le\MmL$.
Since $L\ge L(k)$, we have
\eq
\Biggl|
\frac{\GSb(\hp_L)^{k}\GS}{\GSb(\hp_L)^{k-1}\GS}
-
\lim_{L'\up\infty}\frac{\bra{\Phi_{\rm gs}^{L'}}(\hp_{L'})^{k}\ket{\Phi_{\rm gs}^{L'}}}{\bra{\Phi_{\rm gs}^{L'}}(\hp_{L'})^{k-1}\ket\Phi_{\rm gs}^{L'}}
\Biggr|
\le\epsilon(\MmL)
\lb{LSp30newB}
\en
With \rlb{LSp26}, this implies
\eq
\Bigl|
\bra{\Xi^{(\MmL)}_L}\hp_L\ket{\Xi^{(\MmL)}_L}-\lim_{L'\up\infty}
\bra{\Xi^{(\MmL)}_{L'}}\hp_{L'}\ket{\Xi^{(\MmL)}_{L'}}
\Bigr|\le O\Bigl(\frac{1}{V}\Bigr)+\epsilon(\MmL),
\en
which further implies
\eq
\limL\bra{\Xi^{(\MmL)}_L}\hp_L\ket{\Xi^{(\MmL)}_L}
=\limM\limL\bra{\Xi^{(M)}_L}\hp_L\ket{\Xi^{(M)}_L}.
\en
The expectation value of $\ho^+$ is treated in exactly the same manner.

We should note that, although the present construction proves the existence of $\MmL$, it does not tell us whether a concrete choice of $\MmL$, say $\MmL=L^2$, is suitable.

\subsection{Symmetry breaking order parameter for SU(2) invariant models}
\label{s:su2}
Let us assume that the model has full SU(2) symmetry, and prove Theorem~\ref{t:KomaTasaki3}, i.e., the inequality $\ms\ge\sqrt{3q_0}$.
We follow the argument in \cite{DysonLiebSimon1978,KomaTasaki1993}, which resembles the proof of Lemma~\ref{L:LSP2}.

Let us define
\eq
\hp':=(\hoo)^2+(\hot)^2+(\hos)^2.
\lb{LSp40}
\en
Exactly as in \rlb{LSp8}, we have 
\eq
\frac{\sbkt{(\hp')^n}}{\sbkt{(\hp')^{n-1}}}\ge \sbkt{\hp'}=3\sbkt{(\hoo)^2}=3q_0,
\lb{LSp41}
\en
where we used the SU(2) invariance.
In what follows we prove that
\eq
\limn\limV\sqrt{\frac{\sbkt{(\hp')^n}}{\sbkt{(\hp')^{n-1}}}}=\ms.
\lb{LSp42}
\en
Then \rlb{LSp41} implies the desired inequality $\ms\ge\sqrt{3q_0}$.

We start from an elementary technical lemma.
For nonnegative integers $\ell$, $m$, and $n$, let
\eq
A_{\ell,m,n}:=\frac{1}{4\pi}\int_{x^2+y^2+z^2=1}\hspace{-8mm}dx\,dy\,dz\,x^{2\ell}\,y^{2m}\,z^{2n}.
\lb{LSp43}
\en

\begin{Le}
Let $k$ be a positive integer.
For any nonnegative  integers $\ell$, $m$, and $n$ such that $\ell+m+n=k$, we have
\eq
\frac{(2k)!}{(2\ell)!\,(2m)!\,(2n)!}\,A_{\ell,m,n}:=\frac{1}{2k+1}\,\frac{k!}{\ell!\,m!\,n!}.
\lb{LSp44}
\en
\end{Le}
\nproof{Proof}
We first claim that, for any $a,b,c\in\bbR$, there is an identity
\eq
\frac{1}{4\pi}\int_{x^2+y^2+z^2=1}\hspace{-8mm}dx\,dy\,dz\,(ax+by+cz)^{2k}=\frac{1}{2k+1}\,(a^2+b^2+c^2)^k.
\lb{LSp45}
\en
Note that, by rotational and scaling symmetry,  it is sufficient to show \rlb{LSp45} for $(a,b,c)=(0,0,1)$.
This is elementary.
By expanding both the sides of  \rlb{LSp45}  in $a$, $b$, and $c$, and comparing the coefficients, we get \rlb{LSp44}.~\qedm

\bigskip

From the SU(2) symmetry, we have
\eqa
\bkt{(\hoo)^{2k}}&=\frac{1}{4\pi}\int_{x^2+y^2+z^2=1}\hspace{-8mm}dx\,dy\,dz\,
\bkt{(x\hoo+y\hot+z\hos)^{2k}}
\nl&=\sumtwo{\ell,m,m\ge0}{(\ell+m+n=k)}A_{\ell,m,n}
\hspace{2mm}
\mathop{{\sum}^{(2\ell,2m,2n)}}_{\hspace{-8mm}\alpha_1,\ldots,\alpha_{2k}=1,2,3}
\bkt{\ho^{(\alpha_1)}\cdots\ho^{(\alpha_{2k})}}.
\lb{LSp46}
\ena
Here when we sum over $\alpha_1,\ldots,\alpha_{2k}=1,2,3$, we make a restriction that the numbers of 1, 2, and 3 among $\alpha_1,\ldots,\alpha_{2k}$ are $2\ell$, $2m$, and $2n$, respectively.
The sum contains $\dfrac{(2k)!}{(2\ell)!\,(2m)!\,(2n)!}$ terms.
We, on the other hand, observe that
\eqa
\bkt{(\hp')^{k}}&=\bbkt{\{(\hoo)^2+(\hot)^2+(\hos)^2\}^k}
\nl&=\sumtwo{\ell,m,m\ge0}{(\ell+m+n=k)}
\hspace{5mm}\mathop{{\sum}^{(\ell,m,n)}}_{\hspace{-5mm}\beta_1,\ldots,\beta_{k}=1,2,3}
\bbkt{(\ho^{(\beta_1)})^2\cdots(\ho^{(\beta_{k})})^2}.
\lb{LSp47}
\ena
Here the numbers of 1, 2, and 3 among $\beta_1,\ldots,\beta_{k}$ are $\ell$, $m$, and $n$, respectively.
The sum contains $\dfrac{k!}{\ell!\,m!\,n!}$ terms.
As before we can rearrange the order of operators and show that
\eq
\abs{\bbkt{\ho^{(\alpha_1)}\cdots\ho^{(\alpha_{2k})}}-\bbkt{(\ho^{(\beta_1)})^2\cdots(\ho^{(\beta_{k})})^2}}=O\Bigl(\frac{1}{V}\Bigr),
\lb{LSp48}
\en
provided that $\ell$, $m$, and $n$ are common.
Thus, by noting the numbers of terms in the sums, we have
\eqa
&\mathop{{\sum}^{(2\ell,2m,2n)}}_{\hspace{-8mm}\alpha_1,\ldots,\alpha_{2k}=1,2,3}
\bkt{\ho^{(\alpha_1)}\cdots\ho^{(\alpha_{2k})}}
\nl&
\quad
=\frac{(2k)!}{(2\ell)!\,(2m)!\,(2n)!}\Bigl(\frac{k!}{\ell!\,m!\,n!}\Bigr)^{-1}
\hspace{3mm}
\mathop{{\sum}^{(\ell,m,n)}}_{\hspace{-5mm}\beta_1,\ldots,\beta_{k}=1,2,3}
\bbkt{(\ho^{(\beta_1)})^2\cdots(\ho^{(\beta_{k})})^2}+O\Bigl(\frac{1}{V}\Bigr).
\lb{LSp49}
\ena
Substituting this into \rlb{LSp46}, and using \rlb{LSp44}, we get
\eqa
\bkt{(\hoo)^{2k}}&=\frac{1}{2k+1}\sumtwo{\ell,m,m\ge0}{(\ell+m+n=k)}
\hspace{5mm}
\mathop{{\sum}^{(\ell,m,n)}}_{\hspace{-5mm}\beta_1,\ldots,\beta_{k}=1,2,3}
\bbkt{(\ho^{(\beta_1)})^2\cdots(\ho^{(\beta_{k})})^2}
+O\Bigl(\frac{1}{V}\Bigr)
\nl&=\frac{1}{2k+1}\,\bkt{(\hp')^{k}}+O\Bigl(\frac{1}{V}\Bigr).
\lb{LSp50}
\ena
We thus find
\eq
\frac{\bkt{(\hoo)^{2k}}}{\bkt{(\hoo)^{2(k-1)}}}=\frac{2k-1}{2k+1}\,\frac{\bkt{(\hp')^{k}}}{\bkt{(\hp')^{k-1}}}+O\Bigl(\frac{1}{V}\Bigr),
\lb{LSp51}
\en
which, along with  \rlb{LSp10}, proves the desired \rlb{LSp42}.

\subsection{Low-lying states}
\label{s:lls}
Finally we shall prove Theorem~\ref{t:LLS}, which shows that $\ket{\Gamma_L^M}$ defined by \rlb{LSHA2} is a low-lying state.
Recall that the corresponding bound \rlb{Gen9}  for the quantum Ising model is proved easily by noting that the energy expectation value can be written compactly using double commutator as in \rlb{Gen5}.
Unfortunately, the proof of Theorem~\ref{t:LLS}, in which we must control ever increasing number of low-lying states, is much more complicated.
Here we shall describe the difficulties and see how they are resolved.

For notational simplicity, we write $\hHL'=\hHL-\EGS$.
Our goal is to bound the energy expectation value
\eq
\bra{\Gamma_L^M}\hHL'\ket{\Gamma_L^M}=
\frac{\bbkt{(\ho^-)^M\hHL'(\ho^+)^M}}{\bbkt{(\ho^-)^M(\ho^+)^M}},
\lb{LSp70A}
\en
where we shall always assume $M>0$.
(The case with $M<0$ can be treated in exactly the same manner.)
The first difficulty is that the numerator in this case cannot be written in terms of double commutator as in \rlb{Gen5}.
Recall that it was essential in \rlb{Gen5} that $\hHL$ is sandwiched by the same self-adjoint operator $\hOL$ from the both sides.

To overcome this difficulty, Koma and Tasaki \cite{KomaTasaki1994} required that the model has extra $\bbZ_2$ symmetry.
But there is a much simpler method which does not require any extra assumptions.\footnote{%
We learned this method from Hosho Katsura.
A similar technique was used by Sannomiya, Katsura, and Nakayama \cite{SannomiyaKatsuraNakayama2017}.
See (28) and (29) of \cite{SannomiyaKatsuraNakayama2017}.
}
Since $\bbkt{(\ho^+)^M\hHL'(\ho^-)^M}\ge0$, we see from \rlb{LSp70A} that
\eqa
\bra{\Gamma_L^M}\hHL'\ket{\Gamma_L^M}&\le
\frac{\bbkt{(\ho^+)^M\hHL'(\ho^-)^M}+\bbkt{(\ho^-)^M\hHL'(\ho^+)^M}}{\bbkt{(\ho^-)^M(\ho^+)^M}}
\nl&
=\frac{\bbkt{[(\ho^+)^M,[\hHL',(\ho^-)^M]]}}{\bbkt{(\ho^-)^M(\ho^+)^M}},
\lb{LSp70}
\ena
where the  expression in terms of the double commutator is obtained by noting that $\hHL'\GS=0$ and $\bra{\Phi_L^{{\rm gs}}}\hHL'=0$.
The double commutator is explicitly written as
\eqa
[(\ho^+)^M,[\hHL',&(\ho^-)^M]]=
\sum_{k=0}^{M-1}
[(\ho^+)^M,(\ho^-)^{M-1-k}[\hHL',\ho^-](\ho^-)^k]
\nl=&
\sum_{k,\ell=0}^{M-1}(\ho^+)^{M-1-\ell}[\ho^+,(\ho^-)^{M-1-k}[\hHL',\ho^-](\ho^-)^k](\ho^+)^\ell
\nl=&
\sum_{k,\ell=0}^{M-1}(\ho^+)^{M-1-\ell}(\ho^-)^{M-1-k}[\ho^+,[\hHL',\ho^-]](\ho^-)^k(\ho^+)^\ell
\nl&+
\sum_{k,\ell=0}^{M-1}\sum_{m=0}^{M-k-2}
(\ho^+)^{M-1-\ell}(\ho^-)^{M-k-2-m}[\ho^+,\ho^-](\ho^-)^m[\hHL',\ho^-](\ho^-)^k(\ho^+)^\ell
\nl&+
\sum_{k,\ell=0}^{M-1}\sum_{n=0}^{k-1}
(\ho^+)^{M-1-\ell}(\ho^-)^{M-1-k}[\hHL',\ho^-](\ho^-)^{k-1-n}[\ho^+,\ho^-](\ho^-)^n(\ho^+)^\ell.
\lb{LSp77}
\ena

One might expect that it suffices to bound all these terms by using operator norms, as we did in \rlb{Gen7} and \rlb{Gen8}, to get the desired bound \rlb{LSHA3}.
But this expectation turns out to be too optimistic.
Let us take a look at the first line in the right-hand side of \rlb{LSp77}.
Exactly as in \rlb{Gen8}, we have
\eq
\norm{[\ho^+,[\hHL',\ho^-]]}\le\frac{4\zeta^2(o_0)^2h_0}{V},
\lb{LSp78}
\en
which indeed contains the desired factor $V^{-1}$.
If we also bound other $\ho^\pm$ by their norms, we can bound the first line of the right-hand side of \rlb{LSp77} by a constant times $M^2(o_0)^{2M}/V$.
The denominator of \rlb{LSp70}, on the other hand, is bounded from below as $\sbkt{(\ho^-)^M(\ho^+)^M}\ge(2q_0)^M$.
Thus we find that
\eq
\abs{\frac{(\text{first line of RHS of \rlb{LSp77}})}{\sbkt{(\ho^-)^M(\ho^+)^M}}}
\le(\text{const.})\rbk{\frac{(o_0)^2}{2q_0}}^M\,\frac{M^2}{V}.
\en
Since it must be that $2q_0<o_0$, we find that the factor $\{(o_0)^2/(2q_0)\}^M$ grows exponentially with $M$.
This means that the upper bound can be useful only when one fixes $M$ and lets the system size $L$ grow.
This is of course meaningful, but not what we really want.
Recall that the desired bound \rlb{LSHA3} allows $M$ to be as large as $\sqrt{V}$ (which we believe to be optimal).

In order to overcome this difficulty and prove an optimal bound, we have to give up using naive estimates in terms of operator norms.
Instead we use ``renormalized'' bounds stated in the following Lemmas.
These bounds do not contain``bare" factors like $(o_0)^{2M}$, and are instead expressed in terms of the expectation value $\sbkt{\hp^M}$.

\begin{Le}\label{l:R1}
For any positive $n$ such that
\eq
\frac{a}{q_0V}n^2\le\frac{1}{4},
\lb{R1cond}
\en
where $a$ is the constant introduced in (A1),
and any $s_1,\ldots,s_{2n}=\pm$ with $\sum_{j=1}^{2n}s_j=0$, one has
\eq
\frac{1}{2}\sbkt{\hp^n}\le\bigl|\sbkt{\ho^{s_1}\ho^{s_2}\cdots\ho^{s_{2n}}}\bigr|\le2\sbkt{\hp^n}.
\lb{R1main}
\en
\end{Le}
By tightening the condition \rlb{R1cond} for $n$, we can also prove for any $\varepsilon>0$ the stronger bound $\bigl|\sbkt{\ho^{s_1}\ho^{s_2}\cdots\ho^{s_{2n}}}-\sbkt{\hp^n}\bigr|\le\varepsilon\sbkt{\hp^n}$ by the same method (although \rlb{R1main} is sufficient for us).
This can be regarded as a ``renormalized'' version of the much cruder bound \rlb{LSp5} in terms of the operator norm.

To state the next bound we need to introduce  a class of operators which generalize the local Hamiltonian $\hh_x$ and  the Hamiltonian density $\hHL/V=V^{-1}\sum_{\LaL}\hh_x$.
For each $x$ we consider an operator $\hg_x$ which acts at most on $\zeta$ sites including $x$, and satisfies $\snorm{\hg_x}\le g_0$ with a constant $g_0$.
We also assume that  $[\hC_L,\hg_x]=\mu\,\hg_x$ with some integer $\mu$ independent of $x$, i.e., 
the operator $\hg_x$ changes the charge $\hC_L$ by $\mu$.
We then define the corresponding spatial average by  $\hg=V^{-1}\sum_{x\in\LaL}\hg_x$.
For example, $\hg_x=[\hh_x,\hOL^-]$ satisfies these conditions with $\mu=-1$.
We then have the following bound.

\begin{Le}\label{l:R2}
Take any $\hg$ as defined above.
For any positive $K$ such that
\eq
\frac{a}{q_0V}\Bigl(\frac{K+1}{2}\Bigr)^2\le\frac{1}{4},\quad\frac{3\zeta o_0}{\sqrt{2q_0}\,V}K\le1,
\lb{R2cond}
\en
any $s_1,\ldots,s_{K}=\pm$ with $\sum_{j=1}^{K}s_j=-\mu$, and any $\ell=1,\ldots,K$, one has
\eq
\bigl|\sbkt{\ho^{s_1}\cdots\ho^{s_\ell}\,\hg\,\ho^{s_{\ell+1}}\cdots\ho^{s_K}}\bigr|\le
3g_0\times
\begin{cases}
\sbkt{\hp^{K/2}}&\text{if $K$ is even},\\
\sqrt{\sbkt{\hp^{(K+1)/2}}\sbkt{\hp^{(K-1)/2}}}&\text{if $K$ is odd}.
\end{cases}
\lb{R2main}
\en
\end{Le}

Let us proceed to bound the right-hand side of \rlb{LSp70} and prove Theorem~\ref{t:LLS}  by assuming these Lemmas.
The lemmas will be proved at the end of the present subsection.
Since the denominator $\sbkt{(\ho^-)^M(\ho^+)^M}$ of the right-hand side of \rlb{LSp70} is nonnegative, it can be readily lower bounded by using \rlb{R1main} as
\eq
\bbkt{(\ho^-)^M(\ho^+)^M}\ge\frac{1}{2}\sbkt{\hp^M},
\lb{oo>p}
\en
provided that $M\le\sqrt{q_0V}/2$.
This proves that the state $\ket{\Gamma_L^M}$ is well defined.

The numerator of the right-hand side of \rlb{LSp70} is controlled by using the  explicit form \rlb{LSp77} of the double commutator.
Lemma~\ref{l:R2} plays a central role here.

First we set $\hg_x=[\hOL^+,[\hh_x,\hOL^-]]$, which gives $\hg=V^{-1}[\hOL^+,[\hHL,\hOL^-]]=V\,[\ho^+,[\hHL',\ho^-]]$.
The required conditions for $\hg_x$  are satisfied with $\mu=0$ and $g_0=4\zeta^2(o_0)^2h_0$.
Then, by using \rlb{R2main} with $K=2(M-1)$, we can bound the expectation value of the summand of the first sum in the right-hand side of \rlb{LSp77} as
\eqa
&\bigl|\sbkt{(\ho^+)^{M-1-\ell}(\ho^-)^{M-1-k}[\ho^+,[\hHL',\ho^-]](\ho^-)^k(\ho^+)^\ell}\bigr|
\nl&=\frac{1}{V}\bigl|\sbkt{(\ho^+)^{M-1-\ell}(\ho^-)^{M-1-k}\,\hg\,(\ho^-)^k(\ho^+)^\ell}\bigr|
\nl&\le\frac{1}{V}3g_0\,\sbkt{\hp^{M-1}}
\le\frac{6\zeta^2(o_0)^2h_0}{q_0}\,\frac{1}{V}\,\sbkt{\hp^{M}},
\lb{S1}
\ena
where we used $\sbkt{\hp^{M-1}}\le\sbkt{\hp^{M}}/(2q_0)$, which is \rlb{LSp8}.

To bound the expectation values of the second and the third sums  in the right-hand side of \rlb{LSp77}, we now set $\hg'_x=[\hh_x,\hOL^-]$, which gives $\hg'=V^{-1}[\hHL,\hOL^-]=[\hHL',\ho^-]$.
The conditions are satisfied with  $\mu=-1$ and  $g_0'=2\zeta o_0h_0$.
In the second and the third sums, we also have the commutator $[\ho^+,\ho^-]$.
From the definition \rlb{LSHA1} and the assumption (A2), we find
\eq
[\ho^+,\ho^-]=2\frac{\hC_L}{V^2}.
\lb{oo}
\en
The treatment of the commutator varies in the cases with $M_L=0$ or $M_L\ne0$, where $\hC_L\GS=M_L\GS$, as we assumed in (A4).

When $M_L=0$, we have $\hC_L\ho^{s_1}\cdots\ho^{s_\ell}\GS=(\sum_{j=1}^{\ell}s_j)\ho^{s_1}\cdots\ho^{s_\ell}\GS$, which means that $[\ho^+,\ho^-]=2\hC_L/V^2$ can be replaced by a constant whose absolute value does not exceed $2M/V^2$.
Thus the expectation value of the summand of the second sum is bounded by using \rlb{R2main} with $K=2M-3$ as
\eqa
&\bigl|\sbkt{(\ho^+)^{M-1-\ell}(\ho^-)^{M-k-2-m}[\ho^+,\ho^-](\ho^-)^m[\hHL',\ho^-](\ho^-)^k(\ho^+)^\ell}\bigr|
\nl&\le\frac{2M}{V^2}\,\bigl|
\sbkt{(\ho^+)^{M-1-\ell}(\ho^-)^{M-k-2}\,\hg'\,(\ho^-)^k(\ho^+)^\ell}
\bigr|
\nl&\le
\frac{2M}{V^2}\,3g'_0\,\sqrt{\sbkt{\hp^{M-1}}\sbkt{\hp^{M-2}}}
\le\frac{12\,\zeta o_0h_0}{(2q_0)^{3/2}}\,\frac{M}{V^2}\,\sbkt{\hp^{M}}.
\lb{S2}
\ena
The terms in the third sum can be bounded by exactly the same quantity.
The condition \rlb{R2cond} requires that $M$ satisfies
\eq
\frac{aM^2}{q_0V}\le\frac{1}{4},\quad
\frac{6\zeta o_0(M-1)}{\sqrt{2q_0}\,V}\le1.
\en
These are clearly satisfied if $M\le C_1\sqrt{V}$ with
\eq
C_1=\max\{\sqrt{q_0/(4a)},\sqrt{\sqrt{2q_0}/(6\zeta o_0)}\}.
\lb{C1choice}
\en

Noting that the first sum in the right-hand side of \rlb{LSp77} contains $M^2$ terms and the second and the third term together contain $M^3$ terms, we sum up \rlb{S1} and \rlb{S2} to get
\eq
\bigl|\sbkt{[(\ho^+)^M,[\hHL',(\ho^-)^M]]}\bigr|
\le
\Bigl(\frac{A}{2}\,\frac{M^2}{V}+\frac{B}{2}\,\frac{M^4}{V^2}\Bigr)\sbkt{\hp^M},
\lb{S3}
\en
with $A=12\zeta^2(o_0)^2h_0/q_0$ and $B=24\zeta o_0h_0(2q_0)^{-3/2}$.
Thus, going back to \rlb{LSp70}, we obtain from \rlb{oo>p} and \rlb{S3} that
\eq
\bra{\Gamma_L^M}\hHL\ket{\Gamma_L^M}-\EGS=\bra{\Gamma_L^M}\hHL'\ket{\Gamma_L^M}
\le A\,\frac{M^2}{V}+B\,\frac{M^4}{V^2}.
\lb{nu0}
\en
Since we require that $M^2/V\le(C_1)^2$, where $C_1$ has been already chosen as \rlb{C1choice}, and we get \rlb{LSHA3} with $C_2=A+B(C_1)^2$.

When $M_L\ne0$ our treatment is much less satisfactory.
When evaluating the second and the third sums  in the right-hand side of \rlb{LSp77}, we simply treat $[\ho^+,\ho^-]=2\hC_L/V^2$ as a constant whose absolute value does not exceed $2a/V$ because of  (A1).
Then proceeding in exactly the same manner as above, we get
\eq
\bra{\Gamma_L^M}\hHL\ket{\Gamma_L^M}-\EGS=\bra{\Gamma_L^M}\hHL'\ket{\Gamma_L^M}
\le A\,\frac{M^2}{V}+aB\,\frac{M^3}{V}\le(A+aB)\frac{M^3}{V},
\lb{nun0}
\en
which proves the bound \rlb{LSHA3new} with $C_3=A+aB$.
This completes the proof of Theorem~\ref{t:LLS}.

In fact we expect that the bound of the form \rlb{nu0} is valid in the case with $M_L\ne0$ as well because terms with different signs should cancel with each other.
Our estimate however is not as precise to take into account such cancelations.

\sproof{Proof of  Lemma~\ref{l:R1}}
We shall prove
\eq
\bigr|\sbkt{\ho^{s_1}\cdots\ho^{s_{2n}}}-\sbkt{\hp^n}\bigr|\le\frac{1}{2}\sbkt{\hp^n},
\lb{oop}
\en
under the same condition as Lemma~\ref{l:R1}.
The desired bound \rlb{R1main} follows from the triangle inequality.

Let $n=1$.
Then we see by definition that
\eq
\bigr|\sbkt{\ho^{\pm}\ho^{\mp}}-\sbkt{\hp}\bigr|=\frac{1}{2}\bigl|\sbkt{[\ho^+,\ho^-]}\bigr|
\le\frac{a}{V}\le\frac{a\sbkt{\hp}}{2q_0V},
\lb{opmpmp}
\en
where we noted that \rlb{oo} and the assumption (A1) imply $\snorm{[\ho^+,\ho^-]}\le2a/V$, and also recalled the basic assumption $\sbkt{\hp}\ge2q_0$.
The right-hand side of \rlb{opmpmp} is clearly bounded by $\sbkt{\hp}/2$ because $a/(q_0V)\le1/4$ by \rlb{R1cond}.

We now assume the main bound \rlb{R1main} for $n-1$, and shall prove \rlb{oop}.
As we noted in the proof of Lemma~\ref{L:LSP1},  $\ho^{s_1}\ldots\ho^{s_{2n}}$ can be rearranged into any other $\ho^{s'_1}\ldots\ho^{s'_{2k}}$ with $\sum_{j=1}^{2n}s'_j=0$ by making at most $n^2$ exchanges of $\ho^+$ and $\ho^-$.
We can thus write
\eq
\sbkt{\ho^{s_1}\ho^{s_2}\cdots\ho^{s_{2n}}}-\sbkt{\ho^{s'_1}\ho^{s'_2}\cdots\ho^{s'_{2n}}}
=\sum\sbkt{\ho\cdots\ho[\ho,\ho]\ho\cdots\ho}.
\en
It is crucial to note that $\ho^{s''_1}\cdots\ho^{s''_\ell}\GS$ with any $s''_1,\ldots,s''_\ell=\pm$ is an eigenstate of $[\ho^+,\ho^-]=2\hC_L/V^2$.
We then find
\eq
\bigl|\sbkt{\ho^{s_1}\ho^{s_2}\cdots\ho^{s_{2n}}}-\sbkt{\ho^{s'_1}\ho^{s'_2}\cdots\ho^{s'_{2n}}}\bigr|
\le\frac{2a}{V}\sum
\bigl|\sbkt{\ho^{s''_1}\ho^{s''_2}\cdots\ho^{s''_{2(n-1)}}}\bigr|,
\en
where there are at most $n^2$ terms in the sum.
By using the main bound \rlb{R1main} for $n-1$, we get
\eq
\bigl|\sbkt{\ho^{s_1}\ho^{s_2}\cdots\ho^{s_{2n}}}-\sbkt{\ho^{s'_1}\ho^{s'_2}\cdots\ho^{s'_{2n}}}\bigr|
\le\frac{4an^2}{V}\sbkt{\hp^{n-1}}\le\frac{2an^2}{q_0V}\sbkt{\hp^{n}},
\en
where we used $\sbkt{\hp^{n-1}}\le\sbkt{\hp^{n}}/(2q_0)$, which is \rlb{LSp8}.
From the assumption \rlb{R1cond}, the right-hand side is bounded by $\sbkt{\hp^{n}}/2$.
Recalling that $\hp=(\ho^+\ho^-+\ho^-\ho^+)/2$, we get the desired \rlb{oop}.~\qedm

\sproof{Proof of  Lemma~\ref{l:R2}}
The bound \rlb{R2main} is also proved by induction.
When $K=0$, the desired bound \rlb{R2main} reduces to $|\sbkt{\hg}|\le3g_0$, which is trivially valid.

Let us prove \rlb{R2main} by assuming it for $K-1$.
We first treat the case where $K$ is even.
If $\ell<K/2$, we have
\eqa
\ho^{s_1}\cdots\ho^{s_\ell}\hg\ho^{s_{\ell+1}}\cdots\ho^{s_K}
&-\ho^{s_1}\cdots\ho^{s_{K/2}}\hg\ho^{s_{(K/2)+1}}\cdots\ho^{s_K}
\nl&=\sum_{m=\ell+1}^{K/2}\ho^{s_1}\cdots\ho^{s_{m-1}}[\ho^{s_m},\hg]\ho^{s_{m+1}}\cdots\ho^{s_K}.
\lb{ogo1}
\ena
Let $\hf^\pm_x=[\hOL^\pm,\hg_x]$ and  $\hf^\pm=V^{-1}\sum_{x\in\LaL}\hf^\pm_x=[\hOL^\pm,\hg]=V[\ho^\pm,\hg]$.
Clearly $\hf^\pm_x$ satisfies the conditions for $\hg_x$  with $\mu$ replaced by $\mu\pm1$ and the constant $f_0=2\zeta o_0g_0$.
Then the inductive assumption implies
\eqa
 \bigl|\sbkt{\ho^{s_1}\cdots\ho^{s_{m-1}}\hf^{s_m}\ho^{s_{m+1}}\cdots\ho^{s_K}}\bigr|
&\le3f_0\sqrt{\sbkt{\hp^{K/2}}\sbkt{\hp^{(K/2)-1}}}
\nl&\le\frac{6\zeta o_0g_0}{\sqrt{2q_0}}\sbkt{\hp^{K/2}},
\ena
where we also used \rlb{LSp8}.
Going back to \rlb{ogo1} and taking the ground state expectation value and then the absolute value, we get
\eqa
\bigl|\langle&\ho^{s_1}\cdots\ho^{s_\ell}\hg\ho^{s_{\ell+1}}\cdots\ho^{s_K}\rangle
-\langle\ho^{s_1}\cdots\ho^{s_{K/2}}\hg\ho^{s_{(K/2)+1}}\cdots\ho^{s_K}\rangle\bigr|
\nl&\le\frac{K}{2}\frac{1}{V}\frac{6\zeta o_0g_0}{\sqrt{2q_0}}\sbkt{\hp^{K/2}}
=
\frac{3\zeta o_0}{\sqrt{2q_0}\,V}Kg_0\sbkt{\hp^{K/2}}\le g_0\sbkt{\hp^{K/2}},
\lb{A}
\ena
where we noted that the sum in \rlb{ogo1} contains at most $K/2$ terms, and used the second condition on $K$ in \rlb{R2cond}.
Clearly the same bound holds when $\ell\ge K/2$.
We then use the Schwarz inequality to find 
\eqa
&\bigl|\langle\ho^{s_1}\cdots\ho^{s_{K/2}}\hg\ho^{s_{(K/2)+1}}\cdots\ho^{s_K}\rangle\bigr|^2
\nl&\le
\langle\ho^{s_1}\cdots\ho^{s_{K/2}}\ho^{-s_{K/2}}\cdots\ho^{-s_1}\rangle
\langle\ho^{-s_K}\cdots\ho^{-s_{(K/2)+1}}\hg^\dagger\hg\ho^{s_{(K/2)+1}}\cdots\ho^{s_K}\rangle
\nl&\le(g_0)^2
\langle\ho^{s_1}\cdots\ho^{s_{K/2}}\ho^{-s_{K/2}}\cdots\ho^{-s_1}\rangle
\langle\ho^{-s_K}\cdots\ho^{-s_{(K/2)+1}}\ho^{s_{(K/2)+1}}\cdots\ho^{s_K}\rangle
\nl&\le\bigl(2g_0\sbkt{\hp^{K/2}}\bigr)^2,
\lb{B}
\ena
where we are allowed to use \rlb{R1main} because of the first condition on $K$ in \rlb{R2cond}. 
From \rlb{A} and \rlb{B}, we get
\eq
\bigl|\langle\ho^{s_1}\cdots\ho^{s_\ell}\hg\ho^{s_{\ell+1}}\cdots\ho^{s_K}\rangle\bigr|
\le3g_0\sbkt{\hp^{K/2}},
\en
which is the desired \rlb{R2main}.

The case with odd $K$ is almost the same.
Let us be sketchy.
Corresponding to \rlb{A}, we now have
\eqa
\bigl|\langle\ho^{s_1}\cdots\ho^{s_\ell}\hg\ho^{s_{\ell+1}}&\cdots\ho^{s_K}\rangle
-\langle\ho^{s_1}\cdots\ho^{s_{(K-1)/2}}\hg\ho^{s_{(K+1)/2}}\cdots\ho^{s_K}\rangle\bigr|
\nl&\le\frac{K-1}{2}\,\frac{1}{V}\,6\,\zeta o_0g_0\sbkt{\hp^{(K-1)/2}}
\nl&\le
\frac{3\zeta o_0}{\sqrt{2q_0}\,V}(K-1)g_0\sqrt{\sbkt{\hp^{(K-1)/2}}\sbkt{\hp^{(K+1)/2}}}
\nl&\le g_0\sqrt{\sbkt{\hp^{(K-1)/2}}\sbkt{\hp^{(K+1)/2}}},
\ena
and corresponding to \rlb{B}, we have
\eq
\bigl|\langle\ho^{s_1}\cdots\ho^{s_{(K-1)/2}}\hg\ho^{s_{(K+1)/2}}\cdots\ho^{s_K}\rangle\bigr|^2
\le4(g_0)^2\sbkt{\hp^{(K-1)/2}}\sbkt{\hp^{(K+1)/2}}.~\qedm
\en

\section{Discussion}
\label{s:discussion}
In the present paper, by extending and improving the results by Horsch and von der Linden and by Koma and Tasaki, we have developed a fully rigorous and almost complete general theory about the relation between LRO and SSB in the ground state of a finite system where continuous symmetry is spontaneously broken.
We have extended Koma and Tasaki's result to show that a ground state with LRO but without SSB must be accompanied by an ever increasing number of low-lying excited states.
More importantly, we studied the state \rlb{LSHA5} obtained by superposing the low-lying states, and have shown that the state breaks the symmetry to the full extent, and also that the order operators have vanishing fluctuation in this state in the infinite volume limit.
These results are important as they essentially verify the conjecture by Koma and Tasaki that the state \rlb{LSHA5} converges to a physically meaningful ``ground state''.

Let us discuss some problems which remain to be understood.

We argued that Theorem~\ref{t:main} shows that the state \rlb{LSHA5} converges to the physical ``ground state'' in the infinite volume limit.
Although this conclusion is quite reasonable, one can demand much more precise conditions to characterize the physical ``ground state''.
For example, instead of only looking at the fluctuation of the order operator, one may require that the fluctuation of any macroscopic quantities should vanish in the infinite volume limit.
Mathematically this means that one needs to prove that the infinite volume limit of the state \rlb{LSHA5} is ergodic.
(See Definition~2.7 of \cite{KomaTasaki1994} for the notion of ergodicity.)
Unfortunately our current technique allows us to control the fluctuation only in the form of \rlb{LSHA11}.
Proving the ergodicity seems to be an extremely difficult task, which probably requires much more concrete assumptions about the models.

We have repeated the argument of Kaplan, Horsch, and von der Linden to state Theorem~\ref{t:KHL2}, which shows that SSB can be triggered by infinitesimally small symmetry breaking field.
Unfortunately the variational proof provides almost no information (except for the lower bound for the symmetry breaking order parameter) about the nature of the ground state obtained in this manner.
It is desirable to have a method for controlling the ground state obtained by symmetry breaking field without using abstract variational principle. 

In order to answer these (and other) open problems about symmetry breaking ``ground states'', one probably should abandon the present general approach, and make use of specific properties concrete models.
See \cite{Koma2018} and references therein for examples of such studies.

\bigskip
{\small
It is a pleasure to thank Hosho Katsura for his essential contribution in the derivation of \rlb{LSp70}, which considerably simplified the proof, and for useful comments.
I also thank 
Tohru Koma and 
Haruki Watanabe
for indispensable discussions and comments which made the present work possible, and Masaki Oshikawa and Masafumi Udagawa for useful discussions.
I finally thank Akinori Tanaka for a careful reading of the manuscript and various useful comments, and Yuhi Tanikawa for an important comment.
The present work was supported by JSPS Grants-in-Aid for Scientific Research no.~16H02211.
}


\end{document}